# Wavelet Variance for Random Fields: an M-Estimation Framework

**Stéphane Guerrier**[†] **and Roberto Molinari**[‡]

**Abstract:** We present a general M-estimation framework for inference on the wavelet variance. This framework generalizes the results on the scale-wise properties of the standard estimator and extends them to deliver the joint asymptotic properties of the estimated wavelet variance vector. Moreover, this is achieved by extending the estimation of the wavelet variance to multidimensional random fields and by stating the necessary conditions for these properties to hold when the size of the wavelet variance vector goes to infinity with the sample size. Finally, these results generally hold when using bounded estimating functions thereby delivering a robust framework for the estimation of this quantity which improves over existing methods both in terms of asymptotic properties and in terms of its finite sample performance. The proposed estimator is investigated in simulation studies and different applications highlighting its good properties.

**Keywords and phrases:** Robust Inference, Intrinsically Stationary Processes, Joint Normality, Analysis of Variance.

## 1. Introduction

The Wavelet Variance (WV) is a quantity that is frequently used in many domains, from finance to the natural sciences, since it analyses the variance of a stochastic process by decomposing it across scales. This provides the basis for an analysis of variance in dependent data settings which is extremely useful given that many natural phenomena can be interpreted from this perspective and different tests can be derived from this quantity. For an overview of the importance of the WV we refer the reader to Percival and Walden (2006). A few examples of more recent uses of the WV can be found in Gallegati (2012) where the WV is employed to study the dynamics of financial markets and in Jia et al. (2015) where this quantity is used to study the behavior of crude oil prices. In medicine and genetics the WV was used in Xie and Krishnan (2013) to detect electroencephalogram seizure and support epilepsy diagnosis while in Sanderson et al. (2015) it is used to characterize ancestry block structures from observed genomic patterns. Moreover, the WV has a widespread use in geophysics as highlighted, for example, in Foufoula-Georgiou and Kumar (2014). In particular, the Haar WV is directly linked to the Allan Variance (AV) which is a quantity that is frequently used in many domains for the scale-based analysis and parametric estimation of stochastic processes for example (see Percival, 2015, and references therein for a more detailed overview). In addition, when dealing with second-order intrinsically stationary processes (see Christensen, 1991), the WV is a quantity that well summarizes the "information" contained in the spectral density function into a reduced number of parameters by taking averages of the latter over octave bands. Using this compact information, Guerrier et al. (2013) proposed the Generalized Method of Wavelet Moments (GMWM) that uses the WV to estimate the parameters for complex time series models which, for example, are often employed in engineering for inertial sensor calibration purposes (see Titterton and Weston, 2004). These advantages are preserved also in the two-dimensional case which was already studied in Mondal and Percival (2012b) where they mention how the two-dimensional WV is used, for example, in texture classification and segmentation problems (Unser, 1995) as well as for the analysis of the thickness in soil (Lark and Webster, 2004).

Considering the widespread use of this quantity in different settings, the goal of this paper is to propose a WV estimator for regularly spaced data with desirable properties. For this purpose, an M-estimation framework is put forward based on Huber's Proposal 2 (HP2), thereby delivering the following advantages:

- It generalizes the standard estimator of WV proposed by Percival (1995) and preserves its scale-wise properties which were studied in Serroukh et al. (2000). Moreover it extends these properties to the case where the number of scales goes to infinity with the sample size and delivers the joint asymptotic properties of this estimator across all scales. Indeed, the joint asymptotic normality of the WV vector had never been clearly stated for a fixed number of scales and it had never been studied when the scales go to infinity for a one-dimensional stochastic process (or for multidimensional random fields). This therefore allows to better understand the properties of all the methods that make use of the WV vector for estimation and inference.
- It allows to obtain robust estimates of the WV if a bounded function is chosen for the proposed M-estimator. This provides a robust estimator that preserves the asymptotic properties mentioned in the previous point and compares favorably in finite samples to the one proposed by Mondal and Percival (2012a) under relatively weaker assumptions.

In addition to the above advantages, we trivially extend the WV to the multidimensional case using the definitions given in Mondal and Percival (2012b) and allow for non-Gaussian random fields. These advantages are important under many aspects. Indeed, the joint asymptotic properties are essential for any approach that uses the WV estimator at different scales for statistical tests or parametric inference. Recent examples include the methods proposed in Fan and Gençay (2010) and in Thon et al. (2015) who respectively use the WV vector to obtain a unit-root test and a test for isotropy in random fields. Another example is given by Gencay and Signori (2015) who use the WV vector to develop a Portmanteau test. Moreover, a robust estimator of the WV and its joint asymptotic properties can deliver a robust framework for wavelet-based inference and, for example, can deliver a robust approach for the parametric estimation of time series and spatial models if used within the GMWM (see Guerrier and Molinari, 2016).

Given the above motivation, this paper is organized as follows. Section 2 introduces the straightforward extension of the WV to multidimensional random fields, thereby giving the definitions that will be used throughout the paper. Section 3 defines the proposed M-estimator, delivers its asymptotic properties and highlights how it relates to existing results. Section 4 presents a series of simulation studies where the robustness properties of the new estimator are shown and finally Section 5 presents three applied examples highlighting the usefulness of the new robust framework for stochastic processes and two-dimensional data analysis. Section 6 concludes.

## 2. Multidimensional Wavelet Variance

A wavelet decomposition can theoretically be applied to multidimensional random fields as emphasized in Kugarajah and Zhang (1995), although in practice this is usually limited to one- and two-dimensional processes (in rare cases also three-dimensional processes). The WV can therefore be obtained from the wavelet coefficients resulting from this multidimensional decomposition and, for example, the two-dimensional extension of the WV was already studied in Mondal and Percival (2012b). In this section, we use the latter to extend the definition of the WV to multidimensional random fields and, for this reason, in the following paragraphs we formalize the notation and definitions used for this extension. To start, let $(X_k)_{k \in \mathcal{K}}$ denote the random field on the integer lattice $\mathcal{Z}^D$, with $D < \infty$ and $k \in \mathcal{K}$ defined further on, and let $K_d$ (with $1 \leq k_d \leq K_d$) be the length of the random field along dimension $d$ ($d = 1, \dots, D$). Given these preliminary definitions, below we provide a list of notations that are going to be used throughout the paper:

- $k \in \mathcal{K} = \{k \in \mathbb{N}_+^D | k_1 \leq K_1, ..., k_D \leq K_D\}$: the coordinates of the points on $\mathbb{Z}^D$ for each observation, with $K_d$ being the maximum number of observations along dimension $d$. Based on this we define $N = |\mathcal{K}| = \prod_{d=1}^{D} K_d$ as the cardinality of the set $\mathcal{K}$ (i.e. the number of observations in the random field). To abbreviate notation, we will from this point denote a random field as $(X_k)$.
- $\delta_{k,k^*}$: the Euclidean distance between two points $k$ and $k^*$.
- $\jmath \in \mathcal{J} \subset \mathcal{J}^* = \{\jmath \in \mathbb{N}_+^D | j_1 \leq J_1, ..., j_D \leq J_D\}$: the sets containing the combination of scales of decomposition in each dimension where $J_d = \lfloor \log_2(K_d) \rfloor - a_d$ and

$$a_d = \log_2 \left( \frac{2K_d(L_1 - 1)}{K_d + 2L_1 - 4} \right) ,$$

  with $L_1$ being the length of the wavelet filter at the first scale. Based on this we define $J = |\mathcal{J}| = \prod_{d=1}^{D} J_d$ as being the cardinality of the set $\mathcal{J}$ (i.e. the number of scales of decomposition).
- $M_{j_d} = K_d - L_{j_d} + 1$: the number of wavelet coefficients issued from a decomposition along dimension $d$ where $L_{j_d}$ represents the length of the wavelet filter at scale $j_d$.

- $k_\jmath \in \mathcal{K}_\jmath = \{k_\jmath \in \mathbb{N}_+^D | k_1 \leq M_{j_1}, ..., k_D \leq M_{j_D}\}$: the coordinates of the points on $\mathbb{Z}^D$ for each wavelet coefficient belonging to the decomposition at the set of scales $\jmath$. Based on this we define $M_\jmath = |\mathcal{K}_\jmath| = \prod_{d=1}^D M_{j_d}$ as the cardinality of the set $\mathcal{K}_\jmath$ (i.e. the number of wavelet coefficients issued from a decomposition at the set of scales $\jmath$).
- $l \in \mathcal{L} = \{l \in \mathbb{N}_+^D | l_1 \leq L_{j_1}, ..., l_D \leq L_{j_D}\}$: the sets containing the combination of positions along a wavelet filter in each dimension.
- $(h_{j_d,l_d})$: the Daubechies wavelet filter for scale $j_d$ which satisfies

$$\sum_{l_d=0}^{L_{j_d}-1} h_{j_d,l_d} = 0 \quad \text{and} \quad \sum_{l_d=0}^{L_{j_d}-1} h_{j_d,l_d}^2 = \frac{1}{2^{j_d}}.$$

- $(W_{k_\jmath,\jmath})$: the vectorized process of wavelet coefficients issued from the $D$-dimensional wavelet filtering at the set of scales $\jmath$ (see further on). These coefficients are placed in descending order according to $\delta_{k,k^*}$.

**Remark 2.1.** *The definition of the subset $\mathcal{J}$ allows the wavelet coefficients at the last scale of decomposition (in a general dimension $d$) to go to infinity as the sample size goes to infinity. Notice that we have $\lim_{K_d \to \infty} a_d = \log_2(2(L_1 - 1))$ for any $L_1 < \infty$ and, in particular, $a_d = 1$ exactly when using the Haar wavelet filter.*

To clarify this notation, let us give the following examples.

**Example 2.1.** *Suppose we have a one-dimensional random field (i.e. a stochastic process or time series) which we observe on a regular grid in time $t = 1, \ldots, T$. In this case the sets would reduce to scalars where $k$ would correspond to $t$ while $N$ would correspond to $T$, with the distance between time points being $\delta_{t,t^*} = \sqrt{(t - t^*)^2}$. The wavelet coefficients would therefore be defined as $(W_{t,\jmath})$, naturally ordered according to $\delta_{t,t^*}$, for $\jmath = 1, \ldots, J$ and, using the Haar wavelet filter, $J = \lfloor \log_2(T) \rfloor - 1$.*

**Example 2.2.** *Suppose we have a two-dimensional random field which we observe on a regular grid with coordinates $k = (k_1, k_2)$, where $k_1 = 1, \ldots, K_1$ are the coordinates in dimension $d = 1$ and $k_2 = 1, \ldots, K_2$ are the coordinates in dimension $d = 2$, which gives a random field of size $N = K_1 K_2$ with $\delta_{k,k^*} = \sqrt{(k_1 - k_1^*)^2 + (k_2 - k_2^*)^2}$. We therefore have $\jmath = (j_1, j_2)$, with $j_1 = 1, \ldots, J_1$ and $j_2 = 1, \ldots, J_2$, where, using the Haar wavelet filter, we have $J_1 = \lfloor \log_2(K_1) \rfloor - 1$ and $J_2 = \lfloor \log_2(K_2) \rfloor - 1$.*

Keeping this notation in mind and considering the results of Kugarajah and Zhang (1995) and Mondal and Percival (2012b), we start by defining the $D$-fold wavelet coefficients as

$$W_{k_\jmath,\jmath} = \sum_{l \in \mathcal{L}} h_{j_1,l_1} \cdots h_{j_D,l_D} X_{k_1 - l_1, \ldots, k_D - l_D}. \tag{2.1}$$

Given this definition it can be noticed that $\mathbb{E}[W_{k_\jmath,\jmath}] = 0$ since we generally assume that $(X_k)$ (or its centered differencing) is a second-order stationary random field. Having defined the wavelet coefficients for the set of scales $\jmath$, we can now define the corresponding WV, $\nu_\jmath^2$, as

$$\nu_\jmath^2 \equiv \text{var}\left(W_{k_\jmath,\jmath}\right)$$

which, along with (2.1), allows to obtain an expression for the model-implied WV $\nu_\jmath^2(\boldsymbol{\theta})$. For this purpose, referring to the notation used in Mondal and Percival (2012b), let $\delta_{\boldsymbol{u}}$ represent the Euclidean norm of a vector $\boldsymbol{u}$ and let $\varphi_{\boldsymbol{\theta}}(\delta_{\boldsymbol{u}})$ denote the theoretical covariance function of the random field $\boldsymbol{F_\theta}$. Then we have

$$\nu_\jmath^2(\boldsymbol{\theta}) = \sum_{l\in\mathcal{L}} h_{j_1,l_1}\cdots h_{j_D,l_D}\varphi_{\boldsymbol{\theta}}(\delta_l). \tag{2.2}$$

To simplify notation, from now on we will denote this quantity as $\nu_\jmath^2$. In the one-dimensional setting, known analytic expressions are available for this theoretical WV based on the Haar wavelet filter by using the results of Zhang (2008). However, if the theoretical covariance function of the random field $(X_k)$ is known, it is possible to obtain analytic expressions for any WV based on (2.2). An unbiased estimator of this quantity was proposed by Percival (1995) and is given by

$$\tilde{\nu}_\jmath^2 = \frac{1}{M_\jmath}\sum_{k_\jmath\in\mathcal{K}_\jmath} W_{k_\jmath,\jmath}^2 \ . \tag{2.3}$$

The theoretical properties of this estimator were studied in Serroukh et al. (2000) in which the conditions for its asymptotic properties are given. However, as highlighted by Mondal and Percival (2012a), this estimator of WV is not robust and can therefore become considerably biased in the presence of outliers or different forms of contamination. In order to deliver a WV estimator that can be robust, let us re-express (2.3) as an M-estimator with $\hat{\nu}_\jmath^2$ being the implicit solution of

$$\sum_{k_\jmath\in\mathcal{K}_\jmath} \psi(W_{k_\jmath,\jmath}, \nu_\jmath^2) = 0, \tag{2.4}$$

where $\psi(\cdot)$ is a score function which can be unbounded or bounded with respect to $(W_{k_\jmath,\jmath})$. The following theorem states the sufficient condition under which the WV estimator has a bounded influence function and is therefore robust.

**Theorem 2.1.** *Assuming that $(W_{k_\jmath,\jmath})$ is a strictly stationary and ergodic process and that contamination in dimension $d$ is independent from the contamination in dimension $d'$, $\forall d \neq d'$, the* IF *of the estimator of WV is bounded if $\psi(\cdot)$ is bounded.*

The proof of Theorem 2.1 follows the standard proofs for these settings and can be found in Appendix C.1. The definition of a multidimensional contamination model and relative IF is a topic of wider investigation (see, for example, Alqallaf et al., 2009) but an intuitive argument based on Theorem 2.1 would suggest that a bounded $\psi(\cdot)$ function would nevertheless be required to ensure robustness also in different multidimensional contamination settings. Given these considerations, the next section proposes a new M-estimator for the WV based on HP2.

## 3. M-Estimation of Wavelet Variance

Following the form of the estimator defined in (2.4), this section generalizes the standard estimator of WV in (2.3) to an M-estimator and, hence, the properties of this new M-estimator are valid also for the standard estimator. The proposed approach, as mentioned earlier, is based on HP2 (Huber, 1981) that was put forward for the estimation of the scale parameter of the residuals in the linear regression framework. Since in general we suppose that $\mathbb{E}[W_{k_j,j}] = 0$, we use HP2 by defining $r_{k_j,j} = W_{k_j,j}/\nu_j$ as the standardized wavelet coefficients, and $\hat{\nu}_j^2$ is defined implicitly as the solution in $\nu_j^2$ of

$$\frac{1}{M_j} \sum_{k_j \in \mathcal{K}_j} \omega^2 \left( r_{k_j,j}; \nu_j^2, c \right) r_{k_j,j}^2 - a(\nu_j^2, c) = 0 \, , \tag{3.1}$$

where $r_{k_j,j}^2$ is the score function, $\omega(\cdot)$ represent the weights given to this score function and $a(\nu_j^2, c)$ is a correction term to ensure Fisher consistency at the unconditional distribution of the wavelet coefficients $(W_{k_j,j})$. The tuning constant $c$ regulates the trade-off between robustness and efficiency, where $c \to \infty$ corresponds to the classical estimator (with $\omega(r_{k_j,j}; \nu_j^2, c) = 1, \forall k_j$, and $a(\nu_j^2, c) = 1$). A discussion about the choice of this constant can be found in Appendix A. The term $a(\nu_j^2, c)$ depends on the distribution of the stationary process which is assumed for the wavelet coefficients $(W_{k_j,j})$ and on the specific weight function $\omega(\cdot)$. The exact analytic form of this term therefore may be complicated to derive depending on the distributional assumption. Nevertheless this problem can be overcome, for example, by using indirect inference which allows this estimator to be adapted to different distributional assumptions for $(W_{k_j,j})$.

**Remark 3.1.** *In the case where the wavelet coefficients are assumed to follow a Gaussian process, then the correction term $a(\nu_j^2, c)$ can be expressed as $a_\psi(c)$, since it only depends on the value of the tuning constant $c$, and can be found explicitly using the results of Dhrymes (2005).*

Having defined the M-estimator of WV in (3.1), let us now define $\nu_0^2$ as being the true WV for a general scale $j$ and list a set of conditions which allows us to derive the asymptotic properties of this estimator:

**(C1)** $\nu_0^2 \in \{x \in \mathbb{R} \,|\, 0 < x < \infty\}$.
**(C2)** $\mathbb{E}[\psi(W_{k_j,j}, \nu_j^2)] = 0$ if and only if $\nu_j^2 = \nu_0^2$.
**(C3)** The spectral density $S_\psi(\cdot)$ of the process $(\psi(W_{k_j,j}, \nu_j^2))$ is strictly positive at zero-frequency (i.e. $S_\psi(0) > 0$)
**(C4)** $(W_{k_j,j})$ is stationary with mixing coefficient $\alpha_{W_j,k_j}$ such that $\sum_{k_j=1}^{\infty} (\alpha_{W_j,k_j})^{\delta/2+\delta} < \infty$ for some $\delta > 0$.
**(C5)** $\mathbb{E}[W_{k_j,j}] = 0$ and $\mathbb{E}[W_{k_j,j}^4] < \infty$.
**(C6)** The function $\psi(W_{k_j,j}, \nu_j^2)$ is Bouligand-differentiable and $|\partial/\partial\nu_j^2 \mathbb{E}[\psi(W_{k_j,j}, \nu_j^2)]| < \infty$.

Condition **(C1)** requires the true WV $\nu_0^2$ to belong to a compact support and is a very mild condition given the results of Huber (1967). However, a strong assumption is made by Condition **(C2)** which requires the WV to be identifiable based on the function $\psi(W_{k_j,j}, \nu_j^2)$ and this is not an easy condition to verify in general.

**Remark 3.2.** *Condition (C2) has been proven for the Huber and Tukey $\psi$-functions under the assumption that the wavelet coefficients $(W_{k_j,j})$ follow a Gaussian process (see Theorem C.1 in Appendix C.2).*

Condition **(C3)** is a standard and mild assumption to make in order to obtain an asymptotic distribution for the estimator. Conditions **(C4)** to **(C6)** allow to use specific results to prove consistency and asymptotic normality. In particular, Conditions **(C4)** and **(C5)** directly refer to the wavelet coefficient process $(W_{k_j,j})$ where the mixing coefficient $\alpha_{W_j,k_j}$ allows to achieve an appropriate rate of convergence of the estimator which is useful for the joint asymptotic properties (given further on) while the conditions on the moments are helpful to prove uniform integrability for the central limit theorem to be applied. The latter condition is directly satisfied, for example, if the random field is second-order stationary and Gaussian.

**Remark 3.3.** *Condition (C4) becomes stronger to assume as the dimension of the random field grows. Indeed, with larger $D$, there could be subsets of $(W_{k_j,j})$ where the dependence structure does not decay since they will be based on the same distance $\delta_{k,k^*}$ of the underlying random field $(X_k)$. This condition is less strong in the one-dimensional setting where the filtering only occurs in one direction therefore avoiding a repeated dependence structure of the stochastic process which instead could occur when adding other dimensions in the decomposition. As a final note to this, if the random field is assumed to be isotropic, this limitation could be removed by simply removing the coefficients based on the same distance $\delta_{k,k^*}$ (or by taking an average over them).*

Finally, Condition **(C6)** allows to use the standard expansions to prove the asymptotic normality of the proposed estimator and to achieve appropriate rates of convergence for the results to hold.

**Remark 3.4.** *Condition (C6) is respected when using the Huber or Tukey $\psi$-functions. The Bouligand differentiability of the Huber $\psi$-function is discussed in Appendix C.3.*

Given these conditions, the scale-wise asymptotic properties of the proposed estimator are given in Theorem 3.1 below.

**Theorem 3.1.** *Under Conditions (C1) to (C6) we have that*

$$\sqrt{M_j}(\hat{\nu}_j^2 - \nu_j^2) \xmapsto[N\to\infty]{\mathcal{D}} \mathcal{N}\left(0, \frac{S_\psi(0)}{m_j^2}\right)$$

*where $S_\psi(0)$ is the spectral density of $\psi(W_{k_j,j}, \nu_j^2)$ at zero-frequency and*

$$m_j = \mathbb{E}\left[-\frac{\partial}{\partial \nu_j^2}\psi(W_{k_j,j}, \nu_j^2)\right].$$

The proof of this theorem can be found in Appendix C.3. Theorem 3.1 generalizes the results of Serroukh et al. (2000) to M-estimators under much the same conditions, given that Condition **(C1)** is implicitly assumed in their work. The main differences lie in Conditions **(C2)**, **(C5)** and **(C6)** where, instead of condition **(C5)**, Serroukh et al. (2000) use the condition that $\mathbb{E}[|W_{k_j,j}|^{4+2\delta}] < \infty$, $\forall\, \delta > 0$. Conditions **(C2)** and **(C6)** are simply needed to generalize their results to M-estimators based on specific $\psi$-functions. A more detailed comparison with existing conditions and results in Serroukh et al. (2000) and Mondal and Percival (2012a) can be found in Section 3.1. For some results in this paper we consider the Huber and Tukey $\psi$-functions, suggesting to use the Tukey $\psi$-function based on its renown high breakdown point and its bias properties compared to the Huber function. To support the latter case, in Appendix B we compare Huber's and Tukey's weights in the way they control the bias induced by different types of contamination on the resulting estimator. This study leads us to conclude that Tukey's weights appear to be more appropriate for overall bias reduction.

The results of Theorem 3.1 apply to the estimator $\hat{\nu}_j^2$ for each scale $j$ thereby delivering the scale-wise asymptotic properties of the proposed M-estimator. However different results, such as the isotropy tests in Thon et al. (2015) or the asymptotic properties of the GMWM, rely heavily on the joint asymptotic properties of the vector of estimated WV $\hat{\boldsymbol{\nu}}$. In the following paragraphs we will therefore give some results on the joint properties of the proposed M-estimator when we let $J \to \infty$ as $N \to \infty$. The cases where we keep $J$ fixed are a special case whose properties are easily derived based on these results. Moreover, the latter case has already been studied in Guerrier et al. (2013) within the specific case of the time series setting under different conditions. For this purpose, let us define $M_{\tilde{j}} = \min_d M_{J_d}$ and

$$M_{j*} = (M_{\tilde{j}})^D \ ,$$

which allow to define $k_j \in \mathcal{K}_{j*} = \{k_j \in \mathbb{N}_+^D | k_1 \leq M_{\tilde{j}}, ..., k_D \leq M_{\tilde{j}}\}$. With these definitions we intend to study the joint asymptotic properties of the proposed estimator based on the scale of decomposition which delivers the fewest wavelet coefficients. With this in mind, let us consider also the following condition.

**(C7)** $J/\sqrt{M_{j*}} \to 0$.

Condition **(C7)** implies that $M_{j*}$ goes to infinity at a faster rate than $J$. This is needed since we are considering $D$-dimensional random fields where it is possible for $N \to \infty$ even when we only have one $K_d \to \infty$. In the latter case, we would therefore have that $M_{j*}$ would not go to infinity even though the sample size does. Hence, if we define $K_{d*}$ as being the smallest of the $K_d$'s, we would need $K_{d*}$ to go to infinity at a rate which is sufficiently faster than the growth of the set $\mathcal{J}$. However, in the case where $J$ is fixed, all that is needed is simply that $K_{d*} \to \infty$.

**Remark 3.5.** *In the one-dimensional setting, such as stochastic processes and time series, Condition* ***(C7)*** *is always verified. In higher-dimensional settings,* *this condition becomes stronger but can still be verified and becomes*

$$\frac{\prod_{d=1}^{D}(\lfloor \log_2(K_d) \rfloor - a_d)}{\sqrt{(K_{d*} - L_{J_{d*}} + 1)^D}} \to 0$$

*which implies that $K_{d*}$ should be sufficiently "close" to the other $K_d$'s and grows at a similar rate. For example, if we define $K_d = K \; \forall d$ (meaning that the number of observations along each dimension are the same), then Condition **(C7)** is respected (see proof in Appendix C.4).*

With the above considerations, let us study the joint asymptotic properties of $\hat{\boldsymbol{\nu}}$ by defining $\|\cdot\|$ as the Euclidean norm and, to simplify notation, $\boldsymbol{\nu} \equiv \boldsymbol{\nu}(\boldsymbol{\theta})$. This allows us to give the following corollary.

**Corollary 3.1.** *Under the conditions of Theorem 3.1 and condition **(C7)** we have that*

$$\|\hat{\boldsymbol{\nu}} - \boldsymbol{\nu}\| \xrightarrow{\mathcal{P}} 0.$$

This corollary therefore states the mean-squared consistency of the estimator $\hat{\boldsymbol{\nu}}$ regardless of whether $J$ is fixed or goes to infinity with the sample size $N$. Before studying the asymptotic distribution of $\hat{\boldsymbol{\nu}}$, let us first consider a different scale-wise WV estimator which we call $\bar{\nu}_j^2$ defined as the implicit solution in $\nu_j^2$ of

$$\frac{1}{M_{j*}} \sum_{k_j \in \mathcal{K}_{j*}} \omega^2\left(r_{k_j,j}; \nu_j^2, c\right) \, r_{k_j,j}^2 - a(\nu_j^2, c) = 0. \tag{3.2}$$

This estimator is exactly the same as the estimator proposed in (3.1) except that is based on fewer observations (i.e. $M_{j*}$ instead of $M_j$) and the reason for defining this estimator will become apparent further on. Let us denote the corresponding estimated vector as $\bar{\boldsymbol{\nu}}$ and give the following condition.

**(C8)** The function $\psi(W_{k_j,j}, \nu_j^2)$ is twice Bouligand-differentiable.

**Remark 3.6.** *The Huber $\psi$-function does not satisfy condition **(C8)**.*

This condition is useful to obtain results for some technical corollaries and lemmas in Appendix C.6 which are followed by the proof of Theorem 3.2 given below.

**Theorem 3.2.** *Let $\bar{\boldsymbol{\nu}}$ be the implicit solution of*

$$\sum_{k_j \in \mathcal{K}_{j*}} \Psi(\mathbf{W}_{k_j}, \boldsymbol{\nu}) = \mathbf{0}$$

*where*

$$\Psi(\mathbf{W}_{k_j}, \boldsymbol{\nu}) = \begin{bmatrix} \psi(W_{k_j,1}, \nu_1^2) \\ \vdots \\ \psi(W_{k_j,J}, \nu_J^2) \end{bmatrix}$$

*and with*

$$\psi(W_{k_J,J},\nu_J^2) = \omega^2(r_{k_J,J},\nu_J^2,c)r_{k_J,J}^2 - a(\nu_J^2,c) \tag{3.3}$$

*defining a time-invariant function of* $\mathbf{W}_{k_J}$*. Then, under the conditions of Corollary 3.1 and condition **(C8)**, the asymptotic distribution of* $\bar{\boldsymbol{\nu}}$ *is given by*

$$\sqrt{M_{J*}}\boldsymbol{s}^T\boldsymbol{\Sigma}^{-1/2}\left(\bar{\boldsymbol{\nu}} - \boldsymbol{\nu}\right) \xmapsto[M_{J*}\to\infty]{\mathcal{D}} \mathcal{N}\left(0,1\right)$$

*where* $\|\boldsymbol{s}\| = 1$*,* $\boldsymbol{\Sigma} = \mathbf{M}^{-1}\mathbf{S}_\psi(\mathbf{0})\mathbf{M}^{-T}$ *is the asymptotic covariance matrix of* $\bar{\boldsymbol{\nu}}$ *with* $\mathbf{S}_\psi(\mathbf{0})$ *being the power spectral density of* $\Psi(\mathbf{W}_{k_J},\boldsymbol{\nu})$ *and*

$$\mathbf{M} = \mathbb{E}\left[-\frac{\partial}{\partial\boldsymbol{\nu}}\Psi(\mathbf{W}_{k_J},\boldsymbol{\nu})\right].$$

Although only valid for a "truncated" version of the estimator $\hat{\boldsymbol{\nu}}$ since it only uses $M_{J*}$ observations for each scale, this theorem delivers an estimator (i.e. $\bar{\boldsymbol{\nu}}$) with adequate joint asymptotic properties which are extremely useful for any method which uses the joint properties of this estimator such as, for example, the GMWM. The extension of this result to the estimator $\hat{\boldsymbol{\nu}}$ holds if the number of scales $J$ remains fixed but is not necessarily straightforward if $J \to \infty$, although it can be done based on the results of Theorem C.2 and Corollary C.3 in Appendix C.7. Indeed, based on these results, the theorem below gives the joint asymptotic normality of the estimator $\hat{\boldsymbol{\nu}}$ in the one-dimensional case (i.e. $D = 1$) when $J \to \infty$ and using the Haar wavelet filter.

**Theorem 3.3.** *Suppose that we have a stochastic process* $(X_k)_{k\in\mathcal{K}\subset\mathbb{N}_+}$ *which respects the conditions in Theorem 3.1. Moreover, using the Haar wavelet filter, let* $J = \log_2(N^\alpha)$ *with* $0 < \alpha < 1/2$ *and* $M_{J*} = N - 2^J + 1$*. Defining* $\|\boldsymbol{s}\| = 1$*, we have that*

$$\sqrt{M_{J*}}\boldsymbol{s}^T\boldsymbol{\Sigma}^{-1/2}\left(\hat{\boldsymbol{\nu}} - \boldsymbol{\nu}\right) \xmapsto[M_{J*}\to\infty]{\mathcal{D}} \mathcal{N}\left(0,1\right).$$

The proof of this Theorem can also be found in Appendix C.7. This result allows to select a value of $J$ which goes to infinity with $N$ and preserves the joint normality of the proposed WV estimator when applied to a stochastic process.

**Remark 3.7.** *It is straightforward to see that the results in Corollary 3.1, Theorem 3.2 and Theorem 3.3 hold for any* $D$*-dimensional field* ($D < \infty$) *when* $J$ *is fixed and assuming that* $M_{J*} \to \infty$*.*

The above results have allowed to generalize the WV estimator to an M-estimator whose scale-wise and joint asymptotic properties have been extended to multidimensional random fields and to cases where $J \to \infty$. Moreover, by choosing a bounded $\psi$-function, the proposed estimator can be made robust thereby allowing for WV-based analysis when the data is contaminated. For example, if the choice of the bounded function were to fall on the Tukey $\psi$-function, the following corollary states the asymptotic properties of the corresponding estimator $\hat{\boldsymbol{\nu}}$ which can be used in many practical settings.

**Corollary 3.2.** *Assuming that $(W_{k_J,J})$ is a stationary Gaussian process as well as $J$ fixed, and under Conditions (C1), (C3), (C4), we have that the estimator $\hat{\boldsymbol{\nu}}$ based on the Tukey $\psi$-function is mean-square consistent and has an asymptotic multivariate normal distribution given in Theorem 3.2.*

Given these properties, the following paragraphs compare the results in the previous paragraphs to those of the main existing estimators of the WV, namely the standard estimator of WV ($\tilde{\nu}_J^2$) and the robust M-estimator ($\check{\nu}_J^2$) proposed by Mondal and Percival (2012a).

### *3.1. Comparison with existing results*

As stated throughout the previous sections, the results of this paper share conditions and extend on previous results for existing estimators of WV. Although these estimators were proposed for time series (and two-dimensional random fields), the conditions used for their properties to hold are roughly the same as those for the estimator proposed in the previous section. Considering this, first of all the estimator in (3.1), based on HP2, corresponds to the standard estimator of WV when the weights $\omega(\cdot)$ are all equal to one, i.e. $\tilde{\nu}_J^2$ is the solution for $\nu_J^2$ of

$$\frac{1}{M_J} \sum_{k_J \in \mathcal{K}_J} \frac{W_{k_J,J}^2}{\nu_J^2} - 1 = 0,$$

which obviously has an explicit solution that corresponds exactly to the estimator in (2.3). Hence, the results presented in Section 3 need to be compared to those given in Serroukh et al. (2000). Moreover, the M-estimator in (3.1) is a robust alternative to the existing M-estimator proposed by Mondal and Percival (2012a) which was developed for Gaussian time series and random fields specifically affected by scale-based contamination and makes use of a log-transformation of the wavelet coefficients to apply standard M-estimation theory for location parameters. More specifically, the wavelet coefficients $(W_{k_J,J})$ are transformed as $Q_{k_J,J} = \log W_{k_J,J}^2$ with location parameter $\mu$. The implicit estimator for the solution point $\mu_0$ is given by

$$T_N = \underset{\mu_0 \in \mathbb{R}}{\operatorname{argmin}} \left| \sum_{k_J \in \mathcal{K}_J} \psi(Q_{k_J,J} - \mu_0) \right|.$$

The estimator $T_N$ is consistent for $\mu_0$ which does not necessarily correspond to the location parameter of interest $\mu$ and, consequently, $T_N$ is corrected for bias and then an inverse transform of $T_N$ is applied to deliver the robust estimator of WV.

These two estimators are therefore the terms of comparison for the new M-estimator and let us first compare our conditions and results with those of Serroukh et al. (2000). Without loss of generality, we will only consider $(X_k)$ as being a stochastic process (i.e. one-dimensional random field). Having stated this, let us define the following settings:

**(S1)** Conditions in Lemma 1 in Serroukh et al. (2000):

(a) The stochastic process $(X_k)$ is a strictly stationary process;

(b) $\mathbb{E}[|X_k|^{4+2\delta}] < \infty$ for some $\delta > 0$;

(c) The process $(X_k)$ has mixing coefficient $\alpha_{X_k} = \mathcal{O}(\rho^k)$ with $k \to \infty$ and $0 < \rho < 1$;

(d) The spectral density of the process $(X_k)$ is strictly positive at zero frequency;

(e) $\jmath < \infty$.

**(S2)** Conditions of Theorem 1 in Serroukh et al. (2000):

(a) The stochastic process $(X_k)$ is a strictly stationary process;

(b) $\mathbb{E}[|W_{k_\jmath,\jmath}|^{4+2\delta}] < \infty$ for some $\delta > 0$;

(c) The process $(W_{k_\jmath,\jmath})$ has mixing coefficient $\alpha_{W_\jmath,k_\jmath}$ such that $\sum_{k_\jmath=1}^{\infty}(\alpha_{W_\jmath,k_\jmath})^{\delta/2+\delta} < \infty$ for some $\delta > 0$;

(d) The spectral density of the process $Z_{k_\jmath,\jmath} = W_{k_\jmath,\jmath} - \mathbb{E}[W_{k_\jmath,\jmath}]$ is strictly positive at zero frequency;

(e) $\jmath < \infty$.

**(S3)** Comparable conditions for the M-estimator $\hat{\nu}_\jmath^2$:

(a) **(C3)** to **(C5)**

(b) $\jmath \in \mathcal{J} = \{2^{1,\dots,J}\}$ where $J \to \infty$ by the definition given in Section 2.

The first two settings, as highlighted, correspond to the conditions given in Serroukh et al. (2000) where Setting **(S1)** gives sufficient (but not necessary) conditions for the conditions in Setting **(S2)** to hold. Hence, we have that Setting **(S1)** implies Setting **(S2)**. On the other hand, Setting **(S3)** gives the conditions for the proposed estimator $\hat{\nu}_\jmath^2$ which can be considered comparable to the other settings. Indeed, Conditions **(C1)** and **(C2)** are implicitly assumed in the other settings while Condition **(C6)** is not needed since the $\psi$-function gives the standard estimator in this comparison. The conditions in Setting **(S2)** and Setting **(S3)** are roughly the same except that in the latter we let $\jmath \to \infty$. Having said this, Setting **(S2)** allows to obtain scale-wise asymptotic normality and mean-square consistency of the estimated vector $\hat{\boldsymbol{\nu}}$ for $\jmath < \infty$. In addition to these results, Setting **(S3)** also allows $\jmath \to \infty$ and adds the results on the joint normality of the estimated vector $\bar{\boldsymbol{\nu}}$ since, for the standard estimator in the one-dimensional setting, Conditions **(C7)** and **(C8)** automatically hold. Moreover, remaining in the domain of stochastic processes and by defining $J = \log_2(N^\alpha)$, $0 < \alpha < 1/2$, Setting **(S3)** also delivers the joint normality of the estimated vector $\hat{\boldsymbol{\nu}}$ based on the Haar wavelet filter.

Let us now compare the proposed estimator with the one proposed in Mondal and Percival (2012a) and, for this purpose, let us give the following settings:

**(S4)** The stochastic process $(X_k)$ is Gaussian.

**(S5)** Conditions in Theorem 1 in Mondal and Percival (2012a):

(a) $(X_k)$ is a stationary stochastic process.

(b) $\mathbb{E}[\psi(Q_{k_j,j} - \tilde{\mu})] = 0$ if and only if $\tilde{\mu} = \mu_0$.

(c) The function $\mathbb{E}[\psi(Q_{k_j,j} - \tilde{\mu})]$ is differentiable and is continuous around $\mu_0$.

(d) The process $(X_k)$ has a square integrable spectral density.

Setting **(S5)** partially overlaps with Setting **(S3)** (e.g. Conditions **(C4)** and **(C5)**) and adds the other conditions stated in this paper (i.e. Conditions **(C2)** and **(C6)**). Condition **(C1)** is implicitly assumed for Mondal and Percival (2012a) as well, while an additional condition for them is given by Setting **(S4)** where they require the random field $(X_k)$ to be Gaussian. Considering these relations, under Settings **(S3)** to **(S5)**, the estimator $T_N$ is asymptotically normally distributed with expectation $\mu_0$. These results are not exactly comparable to the scale-wise properties of the estimator $\hat{\nu}_j^2$ since they refer to different quantities of interest. Nevertheless, for their respective quantities and with the exception of the condition that $(X_k)$ be a Gaussian process, these estimators basically share the same type of conditions to obtain scale-wise asymptotic normality and mean-square consistency, assuming the convergence rate of the estimator proposed by Mondal and Percival (2012a) remains unaffected by the bias correction and inverse transform. If however the Gaussian assumption were kept and the Huber and Tukey $\psi$-functions were considered for the estimator $\hat{\nu}_j^2$, then Setting **(S5)** would not be needed anymore for $\hat{\nu}_j^2$ since Conditions **(C2)** and **(C6)** would automatically hold (see Appendices C.2 and C.3). Moreover, as in the comparison with the settings established in Serroukh et al. (2000), the properties of the proposed estimator $\hat{\nu}_j^2$ can be extended to obtain joint asymptotic normality of the vector estimators $\bar{\boldsymbol{\nu}}$ and $\hat{\boldsymbol{\nu}}$. Figure 1 visually summarizes the statements made in this section, attempting to clarify the links between conditions and results of these three estimators.

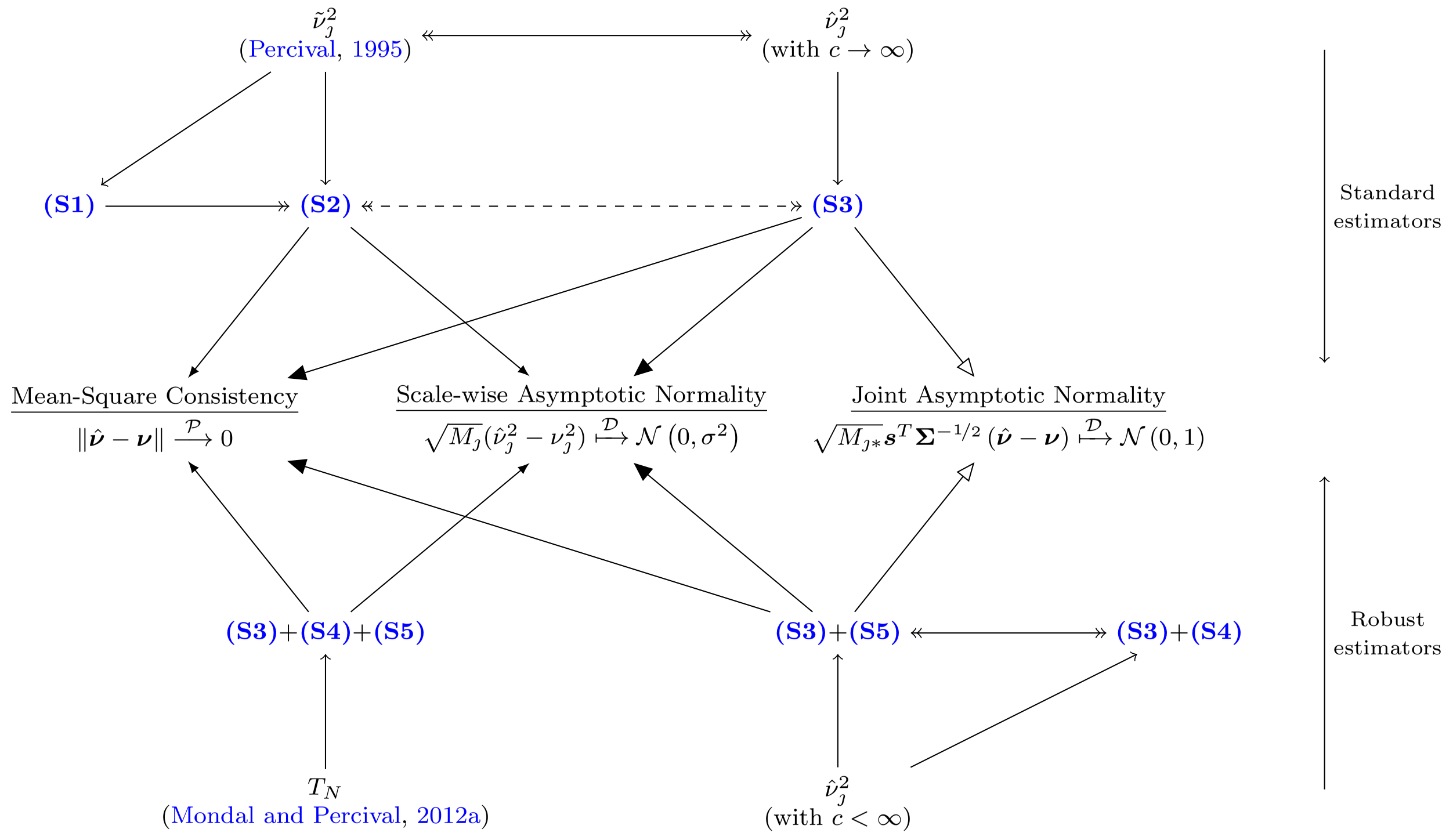

Fig 1: Summary of conditions and asymptotic properties for the standard estimator $\tilde{\nu}_j^2$, the robust estimator $T_N$ and the proposed estimator $\tilde{\nu}_j^2$. Normal arrows indicate links with conditions; double-headed arrows denote equivalence or implication; small full-headed arrow denotes the consequent property under the condition that $j \in J < \infty$; large full-headed arrow denotes the consequent property with $J < \lfloor \log_2(T) \rfloor$; empty-headed arrow denotes the consequent property under the condition that $J < \lfloor \log_2(\sqrt{T}) \rfloor$. The dashed lines denote approximate equivalence.

As mentioned in the introduction, the results of this section allow to obtain more clear and milder conditions for the asymptotic properties of different tests and estimation procedures to hold. As an example, the next section briefly discusses how these results can be important to determine the asymptotic properties of the GMWM in order to carry out parametric estimation and inference for random fields.

### *3.2. Asymptotic Properties of the GMWM*

As highlighted in the introduction, the GMWM delivers a method to estimate a class of basic time series models as well as many linear state-space and structural models. To do so, the GMWM uses the vector of WV as an auxiliary parameter and is defined as

$$\hat{\boldsymbol{\theta}} = \underset{\boldsymbol{\theta} \in \boldsymbol{\Theta}}{\text{argmin}}\, (\hat{\boldsymbol{\nu}} - \boldsymbol{\nu}(\boldsymbol{\theta}))^T \boldsymbol{\Omega} (\hat{\boldsymbol{\nu}} - \boldsymbol{\nu}(\boldsymbol{\theta})) \tag{3.4}$$

where $\boldsymbol{\theta}$ represents the vector of parameters defining the model $F_{\boldsymbol{\theta}}$, $\boldsymbol{\nu}(\boldsymbol{\theta}) = [\nu_j^2(\boldsymbol{\theta})]_{j=1,\dots,J}$ represents the vector of WV implied by the model of interest and $\boldsymbol{\Omega}$ is a weighting matrix chosen in a suitable manner (see Guerrier et al., 2013).

The conditions for consistency and asymptotic normality of the GMWM are roughly the same as those of generalized method of moment estimators. Among these different conditions that are given in Guerrier et al. (2013), two necessary conditions are the following:

**(A1)** $\|\hat{\boldsymbol{\nu}} - \boldsymbol{\nu}\| \xrightarrow{\mathcal{P}} 0$.

**(A2)** $\sqrt{M_{j*}}\boldsymbol{s}^T\boldsymbol{\Sigma}^{-1/2}(\hat{\boldsymbol{\nu}} - \boldsymbol{\nu}) \xrightarrow[M_{j*}\to\infty]{\mathcal{D}} \mathcal{N}(0,1)$, where $\|\boldsymbol{s}\| = 1$.

These conditions are verified under Corollary 3.1, Theorem 3.2 and Theorem 3.3 given in the previous sections. Moreover, these conditions are verified also when the number of scales $J$ goes to infinity (as highlighted in Guerrier and Molinari, 2016) allowing the GMWM to make use of additional information from the sample and continue to benefit from its asymptotic properties for statistical inference on random fields. The results of the previous sections are therefore important to obtain the asymptotic properties of the GMWM estimator and can be used for all other inference procedures which make use of the WV vector such as, for example, Portmanteau and isotropy tests.

## 4. Simulation Studies

To highlight the advantages of the new M-estimator of WV, in this section we briefly present the results of some simulation studies based on different Gaussian time series and spatial models and different contamination settings. The simulated time series are of length $T = 1,000$ while the spatial data is of size $N = 30 \times 30 = 900$. We use the Tukey $\psi$-function for the proposed estimator of WV $\hat{\boldsymbol{\nu}}$ whose asymptotic properties are given in Corollary 3.2 and denote it as RWV, thereby comparing its finite sample performance with:

- $\tilde{\boldsymbol{\nu}}$: the WV estimator from Percival (1995) for time series and Mondal and Percival (2012b) for spatial data (CL);
- $\check{\boldsymbol{\nu}}$: the robust WV estimator proposed by Mondal and Percival (2012a) (MP).

For the proposed M-estimator we choose a tuning constant $c = 4.97$ which delivers 60% efficiency with respect to the standard estimator under the Gaussian assumption. The level of efficiency for $\hat{\boldsymbol{\nu}}$ therefore aims for a higher degree of robustness which approaches the degree provided by the median-type estimator used for $\check{\boldsymbol{\nu}}$ in the simulations in Mondal and Percival (2012a) hence allowing for a fairer comparison between their estimator and the one proposed in this paper. The performance of the estimators is measured by a relative and robust version of the root mean squared error (RMSE*) defined as follows

$$\text{RMSE*} = \sqrt{\text{med}\left(\frac{\bar{\nu}_i - \nu_{i,0}}{\nu_{i,0}}\right)^2 + \text{mad}\left(\frac{\bar{\nu}_i}{\nu_{i,0}}\right)^2}$$

where, for this section, $\bar{\nu}_i$ represents the $i^{th}$ element of a generic vector of estimated WV and $\nu_{i,0}$ is the true WV implied by the model.

To compare the above estimators, we investigate their performance on the first 4 scales of WV ($\jmath = 1, \dots, 4$) in the time series setting and, considering that the spatial models are isotropic, on some elements of the first two columns of the lower diagonal matrix of WV (i.e. $\jmath = (1,1), (1,2), (1,3), (2,2), (3,2)$). The reason for this is that contamination becomes less important at the last scales of decomposition where the length of the filters reduce the effect of outliers by downweighting them with many more uncontaminated observations. In fact, the standard estimator is comparable to the robust estimators at the last scales and this is a phenomenon which is also seen in Mondal and Percival (2012a).

Concerning the contamination settings, different portions of contamination $\epsilon$ were used as well as different types of contamination processes. We use $\sigma_\epsilon^2$ to denote the variance of the zero-mean contaminating observations which were added to the original observations while $\mu_{\epsilon_i}$ denotes the $i^{th}$ mean-shift in the level when using level-shift contamination. With these definitions, the time series models used for the simulations and their relative contamination settings are as follows:

- **AR(1)**: a zero-mean first-order autoregressive model with parameter vector $[\rho_1 \ \upsilon^2]^T = [0.9 \ 1]^T$, scale-based contamination at scale $\jmath = 3$, $\epsilon = 0.01$ and $\sigma_\epsilon^2 = 100$;
- **AR(2)**: a zero-mean second-order autoregressive model with parameter vector $[\rho_1 \ \rho_2 \ \upsilon^2]^T = [0.5 \ -0.3 \ 1]^T$, isolated outliers, $\epsilon = 0.05$ and $\sigma_\epsilon^2 = 9$;
- **ARMA(1,2)**: a zero-mean autoregressive-moving average model with parameter vector $[\rho \ \varrho_1 \ \varrho_2 \ \upsilon^2]^T = [0.5 \ -0.1 \ 0.5 \ 1]^T$, and level-shift contamination with $\epsilon = 0.05$, $\mu_{\epsilon_1} = 5$ and $\mu_{\epsilon_2} = -3$;
- **ARMA(3,1)**: a zero-mean autoregressive-moving average model with parameter vector $[\rho_1 \ \rho_2 \ \rho_3 \ \varrho_1 \ \upsilon^2]^T = [0.7 \ 0.3 \ -0.2 \ 0.5 \ 2]^T$, patchy outliers,

$\epsilon = 0.01$ and $\sigma_\epsilon^2 = 100$;

- **SSM**: a state-space model $(X_t)$ interpreted as a composite (latent) process in certain engineering applications. This model is defined as

$$\begin{aligned} Y_t^{(i)} &= \rho_{(i)} Y_{t-1}^{(i)} + W_t^{(i)} \\ & W_t^{(i)} \overset{\text{iid}}{\sim} \mathcal{N}(0, \upsilon_{(i)}^2) \\ X_t &= \sum_{i=1}^{2} Y_t^{(i)} + Z_t, \\ & Z_t \overset{\text{iid}}{\sim} \mathcal{N}(0, \sigma^2) \end{aligned}$$

with parameter vector

$$[\rho_{(1)}\ \upsilon_{(1)}^2\ \rho_{(2)}\ \upsilon_{(2)}^2\ \sigma^2]^T = [0.99\ 0.1\ 0.6\ 2\ 3]^T,$$

isolated outliers, $\epsilon = 0.05$ and $\sigma_\epsilon^2 = 9$.

Finally, the spatial models used in the simulations and their relative contamination settings are as follows:

- **Exp(1)**: a zero-mean Exponential model with parameter vector $[\phi\ \sigma^2]^T = [2\ 1]^T$ and level-shift contamination with $\epsilon = 0.05$, $\mu_{\epsilon_1} = 5$ and $\mu_{\epsilon_2} = -3$;
- **Exp(2)**: a sum of two zero-mean Exponential models with parameter vector $[\phi_1\ \sigma_1^2\ \phi_2\ \sigma_2^2]^T = [2\ 1\ 1.5\ 1]^T$, isolated outliers, $\epsilon = 0.01$ and $\sigma_\epsilon^2 = 100$;
- **Gauss(1)**: a zero-mean Gaussian model with parameter vector $[\phi\ \sigma^2]^T = [2\ 1]^T$, patchy outliers, $\epsilon = 0.01$ and $\sigma_\epsilon^2 = 100$;
- **Gauss(2)**: a sum of two zero-mean Gaussian models with parameter vector $[\phi_1\ \sigma_1^2\ \phi_2\ \sigma_2^2]^T = [2\ 1\ 1.5\ 1]^T$, isolated outliers, $\epsilon = 0.05$ and $\sigma_\epsilon^2 = 9$;

The results of the simulations for these models can be seen in Figures 2 and 3 where the RMSE* is represented for each scale and the same estimator is connected through a line across scales as a visual support. Figure 2 shows the RMSE* for the estimators based on the time series models while Figure 3 shows it for the spatial models. Within each figure, the top row shows the RMSE* in the uncontaminated setting while the bottom row shows it for the contaminated setting.

The simulations highlight how the estimators all perform better at the first scales and, as expected, gradually become less precise as the scale increases given the lower number of wavelet coefficients. More specifically, however, it shows how the proposed WV estimator RWV is always the best alternative to the classical estimator CL in an uncontaminated setting whereas it is also the best estimator overall when the observed process is contaminated. Indeed, although the MP estimator was proposed for scale-based contamination as for the **AR(1)** simulation, the proposed estimator RWV generally performs better compared to the MP estimator no matter the contamination setting. In addition, the loss of efficiency in the uncontaminated settings with respect to the standard estimator is often visibly small, confirming that the proposed approach provides a generally reliable robust estimator of WV.

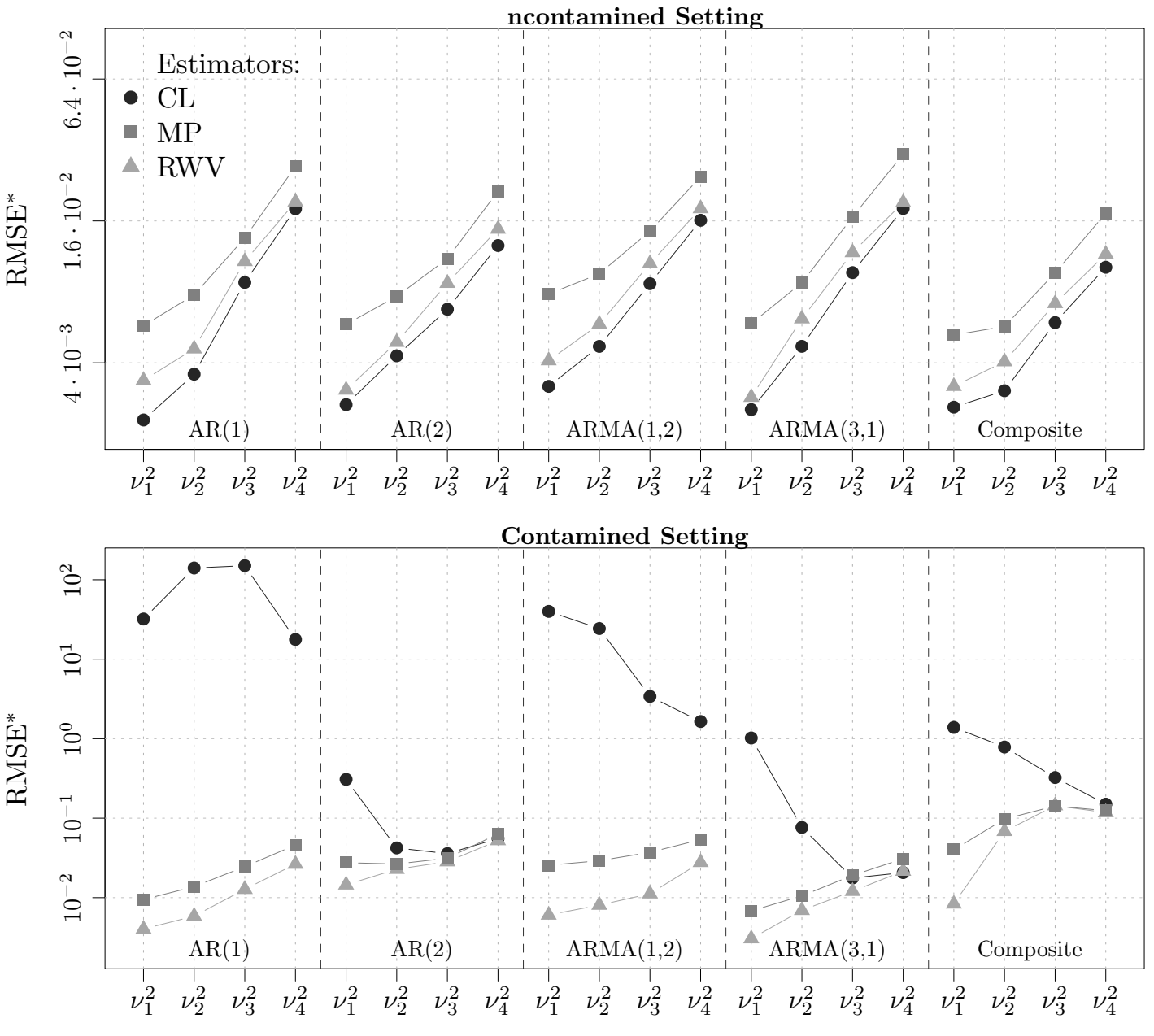

Fig 2: RMSE* of the three WV estimators for the time series models.

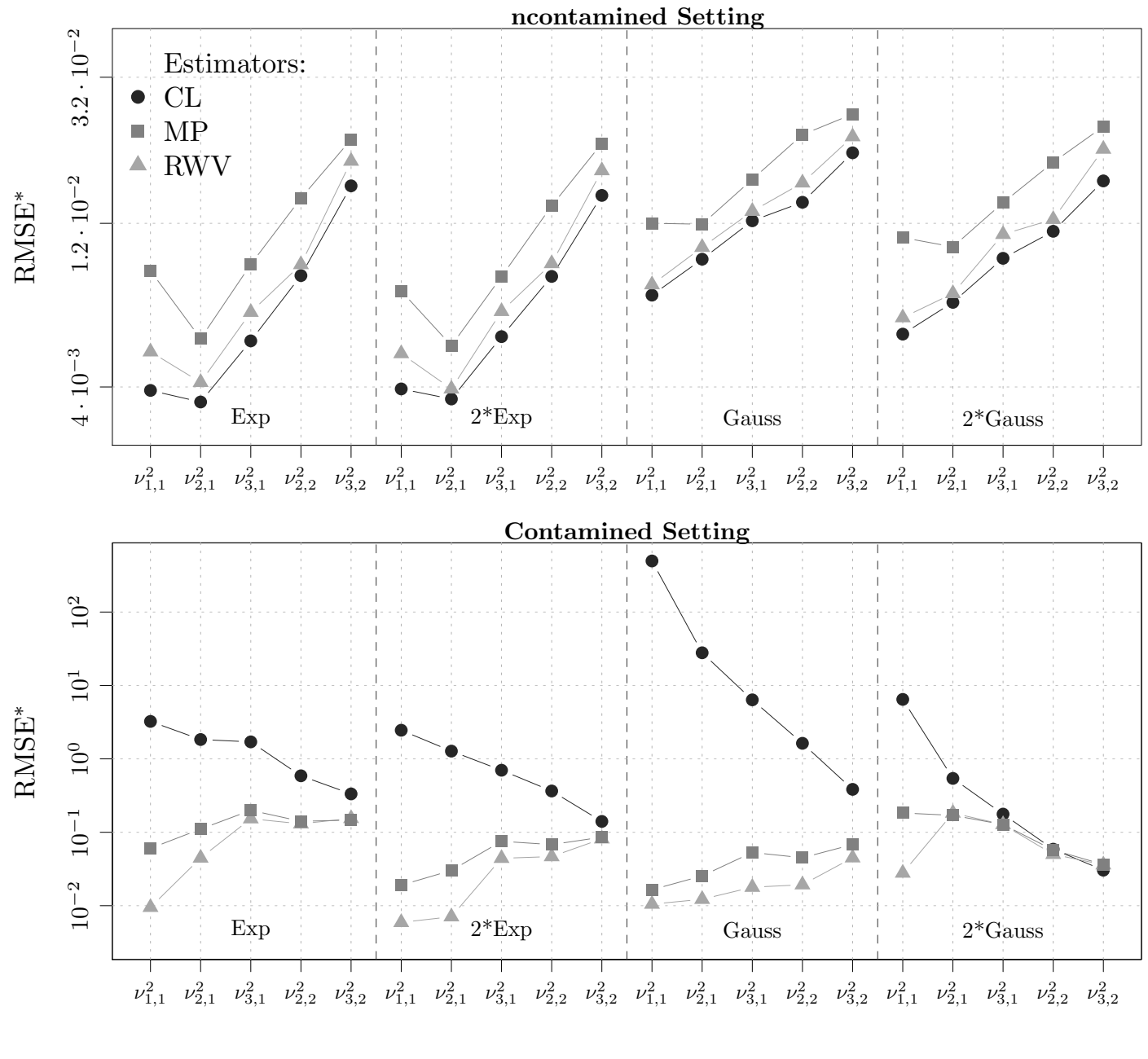

Fig 3: RMSE* of the three WV estimators for the spatial models.

## 5. Applications

In this section we briefly analyse two datasets. The first is the *surface albedo* data which was used as an example in Serroukh et al. (2000) where a one-dimensional WV analysis is employed, while the second one represents a *digital elevation model* where the WV is used to measure the local roughness of this model which is made available through the `raster` package in the `R` statistical software. A third example is given in Appendix D and is taken from Mondal and Percival (2012b) where a two-dimensional random field WV analysis is carried out for the purposes of interpreting and modelling *cloud images*. In all cases, the following paragraphs highlight how a robust approach using the proposed estimator $\hat{\boldsymbol{\nu}}$ is useful to support the overall analysis or reconsider the scientific interpretation of the outputs.

### 5.1. Surface Albedo

Here we revisit the data which was analyzed in Serroukh et al. (2000) as an example of how the WV can be important to determine the quality of a satellite instrument called the "Advanced Very High Resolution Radiometer" (AVHHR) in detecting variability in the ice. In this specific example the data is taken from the measurement of the proportion of incident light reflected (i.e. albedo) of the spring pack ice in the Beaufort Sea and is shown in Figure 4 along with the histograms of the first and second differences of the data.

As highlighted in Serroukh et al. (2000) as well, it can be seen from Figure 4 that there appear to be isolated spikes of low brightness in the data which, according to the authors, can correspond to irregular cracks in the ice or even to surfaces of water. Due to these isolated spikes the authors assume a non-Gaussian process, although they underline that otherwise the series looks rather homogeneous and that the wavelet coefficients appear to be stationary with the exception of some isolated bursts. Indeed, the histograms for the first two backward differences of the data can appear to have a roughly Gaussian distribution if not for the isolated bursts which considerably widen the tails of the distribution. For this reason, we decide to perform a robust WV analysis based on the proposed estimator $\hat{\boldsymbol{\nu}}$ to understand the impact of the isolated bursts on the standard estimator if we assume the wavelet coefficients to follow a Gaussian process. Figure 5 compares the standard estimator of WV with the robust estimator $\hat{\boldsymbol{\nu}}$ for scales $\jmath = 1, \dots, 10$ (which correspond to distances in metres) along with their respective confidence intervals.

It can be observed how the two estimates of WV differ significantly until the $8^{th}$ scale where the standard estimator varies over these scales while the robust estimator remains roughly constant. This agrees with the interpretation given in Serroukh et al. (2000) where the non-Gaussian confidence intervals were used to understand if any conclusions could be drawn on the quality of the AVHHR. In fact, referring to the non-Gaussian confidence intervals in their paper, the authors state that if the true WV were close to the minimum of these confidence intervals up to 800m ($6^{th}$ scale) and close to the maximum of these confidence intervals between 1,600m and 3,200m ($7^{th}$ and $8^{th}$ scales), the AVHHR "*would resolve much of the significant variation*" thereby indicating that the device is of good quality. The robust WV closely follows this interpretation and therefore appears to be a good measure of quality of this device in the presence of isolated bursts in brightness which are frequent in these applications.

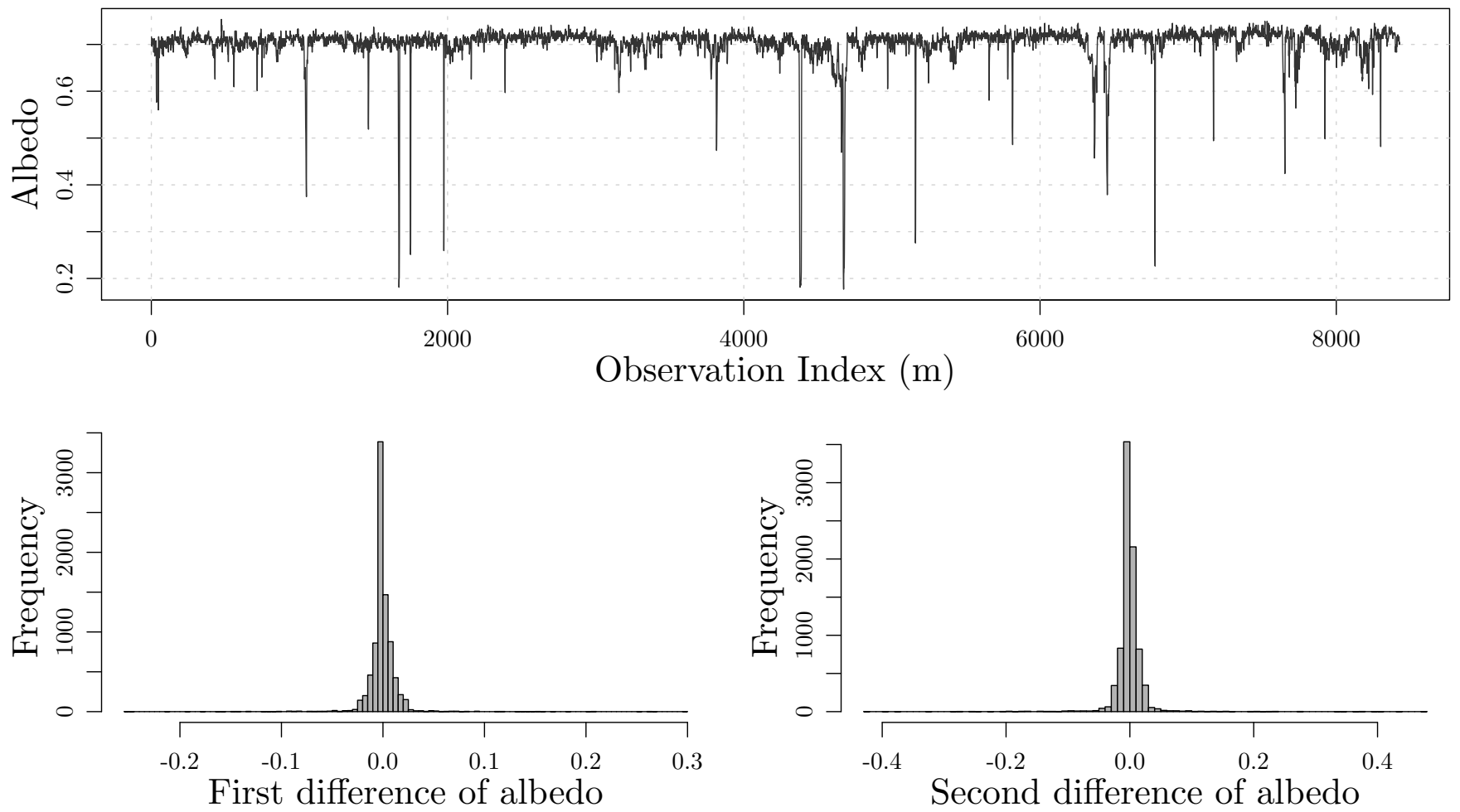

Fig 4: Top: Surface albedo data. Bottom: Histogram of the first backward difference of the surface albedo data (left); Histogram of the second backward difference of the surface albedo data (right).

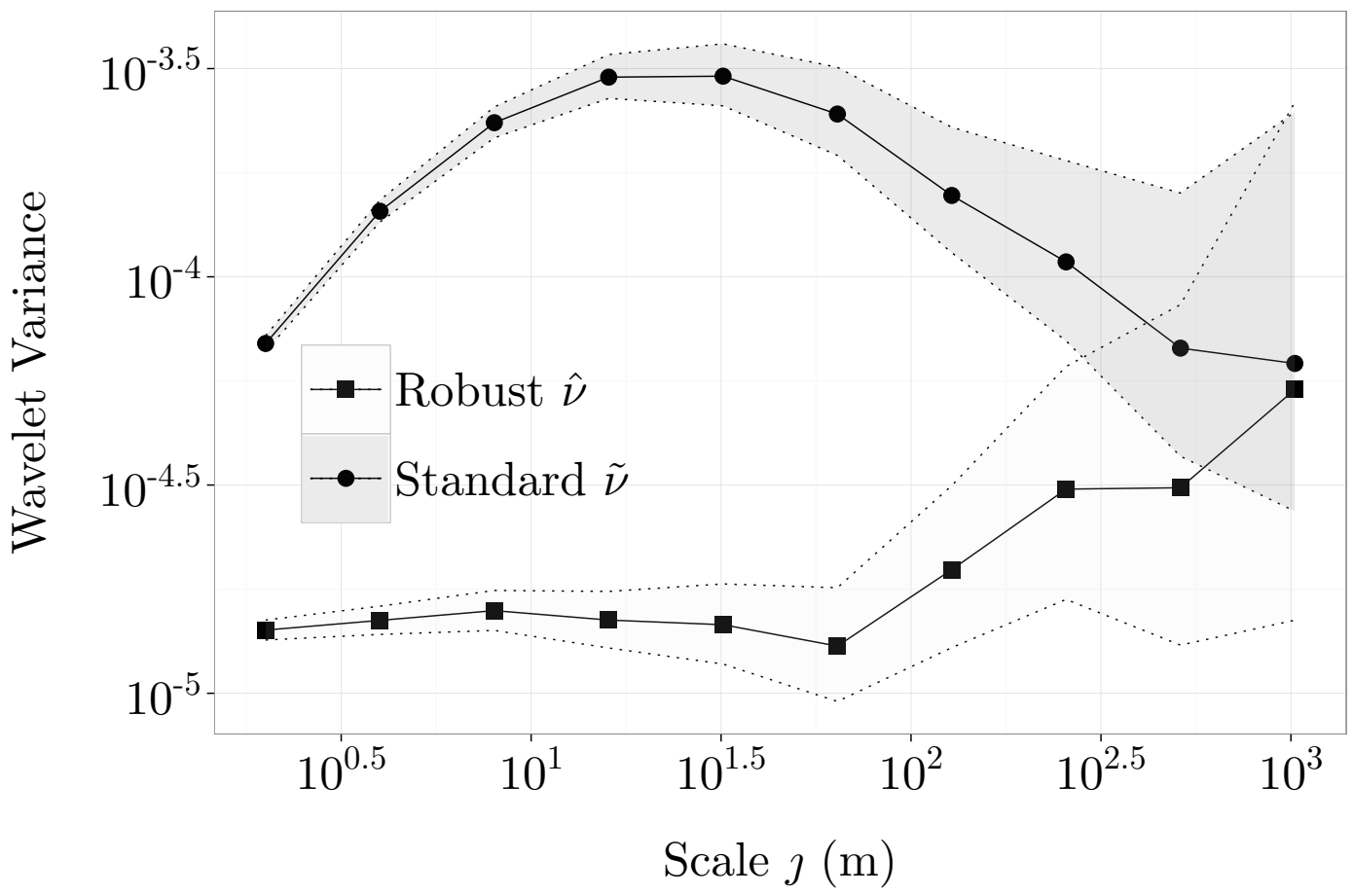

Fig 5: Log-log plot of the estimated standard WV and robust WV on the surface albedo data.

### *5.2. Digital Elevation Models*

Digital Elevation Models (DEM) are digital maps of terrain elevation data and are increasingly being obtained through remote sensing techniques where, for example, radar satellite-based systems are used to generate digital images of surfaces. There is a considerable focus on the quality of DEMs (see for example, Weng, 2002; Wechsler, 2007) since this determines how well these models capture the morphological details of a landscape and to what degree the radar measurements are well approximating the true features of surfaces. The WV is one of the main quantities used to assess the local roughness of the DEM which, if detected, can entail quality improvement techniques such as interpolation thereby improving the geological understanding of the surface (see for example Datcu et al., 1997; Lloyd, 2010). In these cases, an analysis of the diagonal elements of the WV matrix issued from a two-dimensional image decomposition can be extremely useful to understand how well the images are being collected and if some techniques must be used to improve the quality of the DEM. For this purpose, here we carry out a WV analysis on parts of a DEM collected over the San Joaquin field site located in California (USA) and made available through the `raster` package in the R statistical software. Figure 6 shows the log-log plots of the diagonal standard and robust WV taken over specific $1,000\text{m}^2$ areas of the DEM along with their confidence intervals. The vast majority of the WV plots show the same pattern and no significant difference between the standard and robust WV, as shown in the left plot in Figure 6. However, there are some areas of the DEM which appear to suffer from some specific roughness as highlighted by the right plot of Figure 6. Indeed, in the latter the standard WV follows the regular pattern but there appears to be a significant difference with the robust WV at the first three diagonal scales indicating that the pattern for some regions is not the same. This would require to focus on these regions and apply some quality improvement techniques, such as interpolation, in order to make the mapping more reliable.

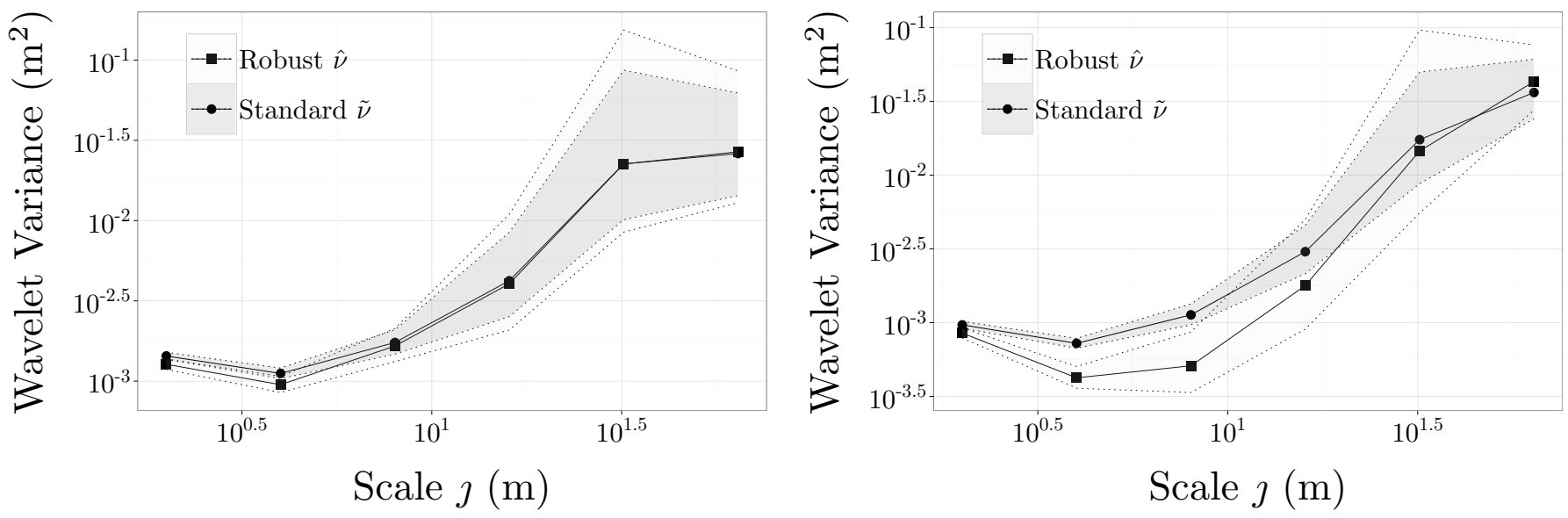

Fig 6: Left: Log-log plot of the standard and robust WV over a $1,000\text{m}^2$ area with coordinates of central point: [4050, 4050]. Right: Log-log plot of the standard and robust WV over a $1,000\text{m}^2$ area with coordinates of central point: [450, 450].

## 6. Conclusion

We have presented a new general framework for the (robust) estimation of the WV in multidimensional random fields. This framework includes and considerably extends the theoretical and asymptotic properties of the standard estimators of WV providing the conditions for the joint properties of this estimator to hold. The latter conditions are especially important for the growing use of the WV for statistical inference, from Portmanteau tests to isotropy tests and parametric estimation of time series and spatial models. Moreover, the M-estimator on which this framework is based can be made robust by choosing a bounded score function ($\psi$-function) allowing to perform WV analysis when the data is contaminated and improving over alternative robust estimators both in terms of clear asymptotic properties as in terms of finite sample performance. The simulation studies and applied examples highlighted how this new framework is statistically sound and is extremely useful for practical purposes when, for example, the data is affected by contamination.

## Appendix A: Choice of tuning constant

Given that Theorem 3.1 provides an expression for the variance of the WV estimator, here we provide a brief discussion on the choice of the tuning constant $c$. The definition of this value is based on the desired level of efficiency compared to the classical estimator and varies according to the chosen $\psi$-function. However, an explicit and intuitive rule for the choice of this constant is available only when considering the process $(W_{k_j,j})$ as Gaussian. In the latter case, the estimator in (3.1) is the result of a minimization under the standard Gaussian assumption (i.e. zero mean and unit variance) and because of this we can obtain expressions for the variance of both the classical estimator and the robust estimator, which we denote as $\sigma_j^2$ and $\tilde{\sigma}_j^2(c)$ respectively. In this setting we see that these expressions depend solely on $c$ and therefore, for a general scale $j$ and defining $\alpha \in [0,1]$ as the desired level of efficiency, a rule to select the tuning constant $c$, given a specific $\psi$-function, is to find the solution in $c$ to the expression

$$\frac{\sigma_j^2}{\tilde{\sigma}_j^2(c)} - \alpha = 0.$$

For example, choosing $\alpha = 0.95$ delivers a tuning constant $c \approx 7.88$ when using the Tukey $\psi$-function and $c \approx 2.38$ when using the Huber $\psi$-function (respectively $c \approx 4.97$ and $c \approx 1.22$ for $\alpha = 0.6$). The choice of the efficiency level is subjective and can be supported by a sensitivity analysis comparing the classical and the robust estimates starting from a low efficiency level (e.g. 0.5).

## Appendix B: Bias Study

We consider the setting where the observations of the random field generate wavelet coefficients issued from a Gaussian process. If one supposes that the Gaussian distribution is at most a fair approximation of the data generating model, then the latter should be considered as belonging to a neighbourhood of the postulated model, say $F_\epsilon$, also called the contaminated distribution. Under $F_\epsilon$, the proposed estimator is hence biased, albeit that the bias is bounded as a result of the bounded $\psi$-functions. The size of the bias depends on the tuning constant $c$ and the form of the $\psi$-function. Hence, by fixing the desired efficiency of the resulting estimator via an appropriate tuning constant $c$, the choice of the $\psi$-function could be seen as a bias-minimization problem.

This bias-minimization problem was considered by Huber (1981) within the "minimax" approach. Hampel et al. (1986) consider the (general) case of a contaminated distribution $F_\epsilon = (1-\epsilon)F_{\boldsymbol{\theta}} + \epsilon H$, $0 \leq \epsilon < 1$. In the present setting, we consider $F_{\boldsymbol{\theta}}$ to be the standard Gaussian model. Let us also write the estimator defined through the function $\psi(\cdot)$ as a functional of the underlying distribution, i.e. $T(F_\epsilon)$ or $T(F_{\boldsymbol{\theta}})$, and we suppose it to be Fisher consistent (i.e. $E_{F_{\boldsymbol{\theta}}}[\psi(r_{k_j,j};c)] = 0$, with $r_{k_j,j}$ being a standard Gaussian variable which represents the standardised wavelet coefficients in the context of this paper. Using a Von Mises expansion of $T(F_\epsilon)$ around $T(F_{\boldsymbol{\theta}})$, the (approximate) asymptotic bias of the estimator is expressed as

$$T(F_\epsilon) - T(F_{\boldsymbol{\theta}}) \approx \epsilon \int \mathrm{IF}(r, \psi, F_{\boldsymbol{\theta}}) dH(r) \tag{B-1}$$

where $\mathrm{IF}(r, \psi, F_{\boldsymbol{\theta}})$ is the IF of the estimator and $r = r_{k_\jmath, \jmath}$ for $\jmath = 1, \dots, J$.

The bias of $T(\cdot)$ under a contaminated distribution $F_\epsilon$ is hence directly proportional to its IF under the contaminating distribution $H$. Therefore, given a desired level of efficiency (i.e. having selected the tuning constant $c$), it is possible to select a $\psi$-function over another one by comparing their (approximate) bias.

The IF of an $M$-estimator is given by

$$\mathrm{IF}(r, \psi, F_{\boldsymbol{\theta}}) = M^{-1}(\psi, F_{\boldsymbol{\theta}}) \psi(r, \boldsymbol{\theta}) \tag{B-2}$$

where $M(\psi, F_{\boldsymbol{\theta}})$ for Fisher consistent $M$-estimators is defined as

$$M(\psi, F_{\boldsymbol{\theta}}) = \int \psi(r, \boldsymbol{\theta}) s^T(r, \boldsymbol{\theta}) dF_{\boldsymbol{\theta}}(r)$$

with $s(r, \boldsymbol{\theta})$ being the score function ($s(r, \boldsymbol{\theta}) = r^2$ in our case). Since $M(\psi, F_{\boldsymbol{\theta}})$ depends on the underlying $\psi$-function and is constant under a postulated model $F_{\boldsymbol{\theta}}$, we will denote it generically as $M_\psi$. Using (B-1) yields the following approximate bias for the proposed estimator

$$B_\psi(H, F_{\boldsymbol{\theta}}) = \epsilon M_\psi^{-1} \int \psi(r, \boldsymbol{\theta}) dH(r). \tag{B-3}$$

If $H = F_{\boldsymbol{\theta}}$, the IF has value 0 and, consequently, so does the bias. As (B-3) shows, given a contaminating distribution $H$, the bias of the proposed estimator ultimately depends on the chosen $\psi$-function. The choice of the $\psi$-function can therefore be made based on the minimization of a risk-function which takes into account expression (B-3). A possible risk-function could simply be

$$tr\left(B_\psi(H, F_{\boldsymbol{\theta}}) B_\psi^T(H, F_{\boldsymbol{\theta}})\right). \tag{B-4}$$

To compare the bias-performance of the Huber and Tukey biweight $\psi$-functions (indexed with $[Hub]$ and $[Bi]$ respectively), we computed (B-4) for $H$ being the dirac distribution with pointmass at $\delta \in [0, 10]$. The tuning constants for the two $\psi$-functions were chosen to guarantee 95% asymptotic efficiency at the normal model, yielding $c_{[Hub]} \cong 2.38$ and $c_{[Bi]} \cong 7.88$.

As the top part of Figure 7 highlights, the risk function of the $\psi_{[Bi]}$-function peaks and descends becoming constant around $c_{[Bi]} \cong 7.88$, whereas the $\psi_{[Hub]}$-function grows and remains constant after $c_{[Hub]} \cong 2.38$. Having approximately the same behavior until around $\delta = 5$ (with the $\psi_{[Bi]}$-function's risk being greater over a small interval), the risk of the $\psi_{[Bi]}$-function is constantly smaller, indicating that the latter function appears to have an overall better performance in terms of risk.

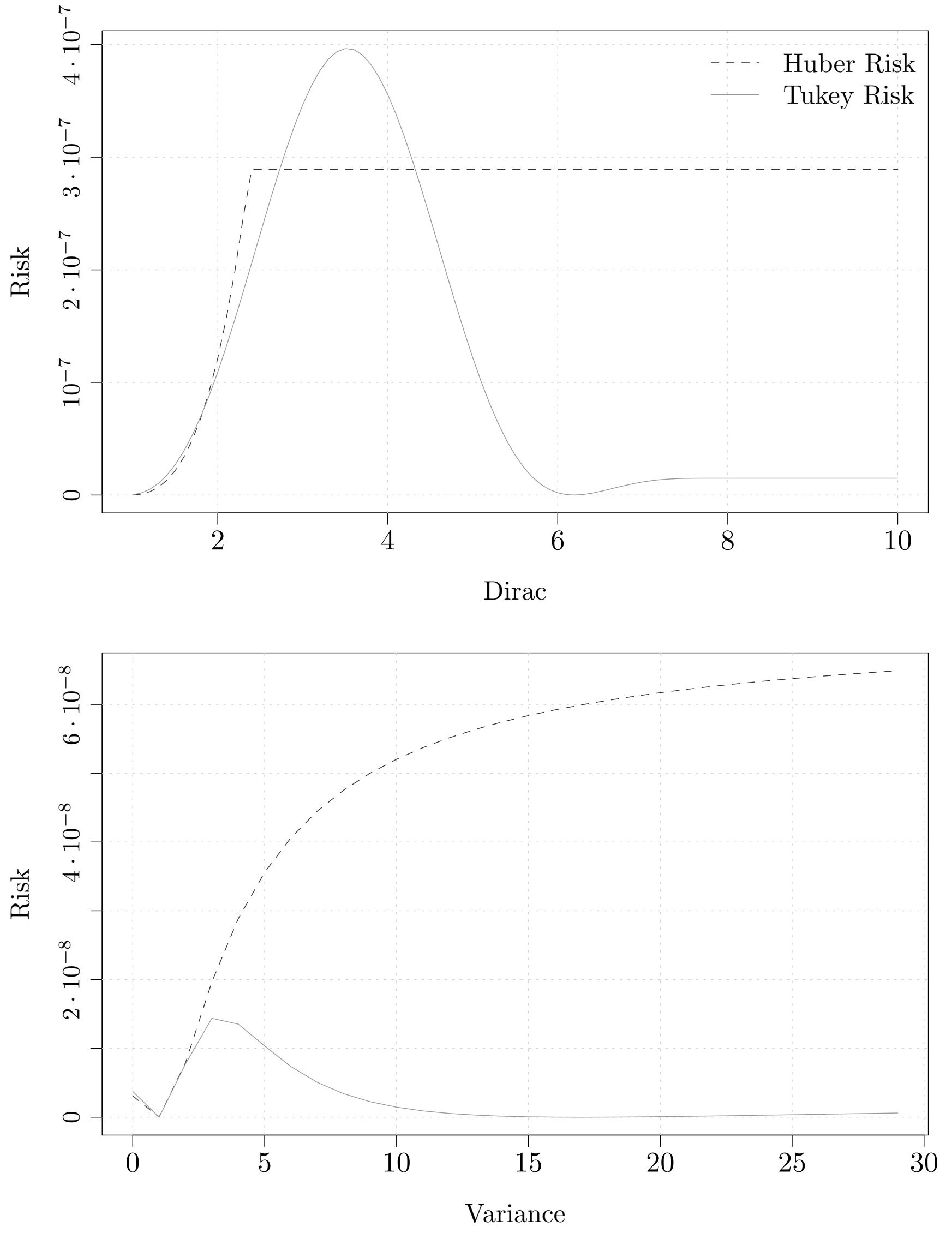

Fig 7: Top: theoretical risk function for a series of Dirac points. Bottom: theoretical risk function for a Gaussian contamination with different scale parameters

In the bottom of Figure 7 the risk function of the $\psi$-functions are computed with $H$ being another zero-mean Gaussian distribution with $\sigma^2 \in (0, 30]$. It can again be seen how the risk of the two functions initially behave in a similar way whereas the $\psi_{[Bi]}$-function's risk becomes constantly smaller compared to the one of the $\psi_{[Hub]}$-function after a given magnitude of contamination. In addition, we can see how both functions deliver unbiased estimators when the contaminating distribution corresponds to the true distribution $F_{\boldsymbol{\theta}}$ (i.e. when the contaminating variance equals 1).

## Appendix C: Proofs

### *C.1. WV Influence Function*

Below is the proof of Theorem 2.1 which follows standard procedures for the derivation of the influence function for M-estimators.

*Proof.* Let us define $\boldsymbol{\epsilon} = [\epsilon_d]_{d=1,\ldots,D}$ as the vector of contamination levels for each dimension $d$ and, under the assumption of Theorem 2.1, let $\varepsilon = \prod_{d=1}^{D} \epsilon_d$. Moreover, let $(X_k)$ be a random field generated by a model in a neighbourhood of $F_{\boldsymbol{\theta}}$, i.e. $F_\varepsilon = (1-\varepsilon)F_{\boldsymbol{\theta}} + \varepsilon\Delta_{\boldsymbol{z}}$, with small $\varepsilon > 0$. $F_\varepsilon$ is the standard contamination model with $\Delta_{\boldsymbol{z}}$ the Dirac function at $\boldsymbol{z} = X_k^\varepsilon$, since the maximal bias on a statistic (estimator) is obtained when the contaminating distribution is a Dirac (see Hampel et al., 1986).

Very generally, let $(x_i), i \in \mathcal{I} \subset \mathbb{N}_+$, represent an observed sequence of a random variable $(X_i)$ following a model $F_{\boldsymbol{\theta}}$. An $M$-estimator is defined as the solution for $\boldsymbol{\lambda}$ of the following equation

$$\sum_{i \in \mathcal{I}} \psi(x_i, \boldsymbol{\lambda}) = 0 \tag{C-1}$$

where $\boldsymbol{\lambda}$ can be $\boldsymbol{\theta}$ or a function of it (i.e. $\boldsymbol{\lambda}(\boldsymbol{\theta})$) and $\psi(\cdot)$ is a function which can be unbounded or bounded. Considering $F_\varepsilon$ as the data generating model, then $\hat{\boldsymbol{\lambda}}(F_\varepsilon)$ (i.e. the estimator written as a functional of $F_\varepsilon$) is implicitely defined in

$$\mathbb{E}_{F_\varepsilon}\left[\psi\left(x_i, \hat{\boldsymbol{\lambda}}(F_\varepsilon)\right)\right] = 0. \tag{C-2}$$

In our case we consider the wavelet coefficients that can also be written as functionals of $F_\varepsilon$, namely $W_{k_\jmath,\jmath}(F_\varepsilon)$, and seek to estimate the WV considering $F_\varepsilon$, i.e.

$$\mathbb{E}_{F_\varepsilon}\left[\psi\left(W^2_{k_\jmath,\jmath}(F_\varepsilon), \hat{\nu}^2_\jmath(F_\varepsilon)\right)\right] = 0. \tag{C-3}$$

Under the assumption of Theorem 2.1 and using the time series adaptation of the IF from Kunsch (1984), the IF of $\hat{\nu}^2_\jmath$ is obtained by taking the Gâteaux derivative of (C-3) with respect to $\varepsilon$ when $\varepsilon \to 0^+$, i.e.

$$\frac{\partial}{\partial \varepsilon}\left[(1-\varepsilon)\mathbb{E}_{F_{\boldsymbol{\theta}}}\left[\psi\left(W^2_{k_j,j}(F_\varepsilon), \hat{\nu}^2_j(F_\varepsilon)\right)\right]\right]_{\varepsilon\downarrow 0} + \frac{\partial}{\partial \varepsilon}\left[\varepsilon\psi\left(W^2_{k_j,j}(F_\varepsilon), \hat{\nu}^2_j(F_\varepsilon)\right)\right]_{\varepsilon\downarrow 0}$$

$$\begin{aligned}
&= -\mathbb{E}_{F_{\boldsymbol{\theta}}}\left[\psi\left(W^2_{k_j,j}(F_{\boldsymbol{\theta}}), \nu^2_j\right)\right] + \\
&\quad \mathbb{E}_{F_{\boldsymbol{\theta}}}\left[\frac{\partial}{\partial W^2_{k_j,j}}\psi\left(W^2_{k_j,j}, \nu^2_j\right)\right]_{W^2_{k_j,j}=W^2_{k_j,j}(F_{\boldsymbol{\theta}})} \frac{\partial}{\partial \varepsilon}\left[W^2_{k_j,j}(F_\varepsilon)\right]_{\varepsilon\downarrow 0} + \\
&\quad \mathbb{E}_{F_{\boldsymbol{\theta}}}\left[\frac{\partial}{\partial \nu^2_j}\psi\left(W^2_{k_j,j}(F_{\boldsymbol{\theta}}), \nu^2_j\right)\right] \frac{\partial}{\partial \varepsilon}\left[\hat{\nu}^2_j(F_\varepsilon)\right]_{\varepsilon\downarrow 0} + \psi\left(W^2_{k_j,j}(\boldsymbol{z}), \nu^2_j\right) \\
&= -\mathbb{E}_{F_{\boldsymbol{\theta}}}\left[\psi\left(W^2_{k_j,j}(F_{\boldsymbol{\theta}}), \nu^2_j\right)\right] + \\
&\quad \mathbb{E}_{F_{\boldsymbol{\theta}}}\left[\frac{\partial}{\partial W^2_{k_j,j}}\psi\left(W^2_{k_j,j}, \nu^2_j\right)\right]_{W^2_{k_j,j}=W^2_{k_j,j}(F_{\boldsymbol{\theta}})} \mathrm{IF}(\boldsymbol{z}, W^2_{k_j,j}, F_{\boldsymbol{\theta}}) + \\
&\quad \mathbb{E}_{F_{\boldsymbol{\theta}}}\left[\frac{\partial}{\partial \nu^2_j}\psi\left(W^2_{k_j,j}(F_{\boldsymbol{\theta}}), \nu^2_j\right)\right] \mathrm{IF}(\boldsymbol{z}, \hat{\nu}^2_j, F_{\boldsymbol{\theta}}) + \psi\left(W^2_{k_j,j}(\boldsymbol{z}, F_{\boldsymbol{\theta}}), \nu^2_j\right) \equiv \boldsymbol{K}
\end{aligned}$$

By the chain rule we have that $\mathrm{IF}(\boldsymbol{z}, \hat{\nu}^2_j, F_{\boldsymbol{\theta}}) \propto \mathrm{IF}(\boldsymbol{z}, W^2_{k_j,j}, F_{\boldsymbol{\theta}})$, therefore

$$\begin{aligned}
\boldsymbol{K} \propto &-\mathbb{E}_{F_{\boldsymbol{\theta}}}\left[\psi\left(W^2_{k_j,j}(F_{\boldsymbol{\theta}}), \nu^2_j\right)\right] + \\
&\mathbb{E}_{F_{\boldsymbol{\theta}}}\left[\frac{\partial}{\partial W^2_{k_j,j}}\psi\left(W^2_{k_j,j}, \nu^2_j\right)\right]_{W^2_{k_j,j}=W^2_{k_j,j}(F_{\boldsymbol{\theta}})} \mathrm{IF}(\boldsymbol{z}, \hat{\nu}^2_j, F_{\boldsymbol{\theta}}) + \\
&\mathbb{E}_{F_{\boldsymbol{\theta}}}\left[\frac{\partial}{\partial \nu^2_j}\psi\left(W^2_{k_j,j}(F_{\boldsymbol{\theta}}), \nu^2_j\right)\right] \mathrm{IF}(\boldsymbol{z}, \hat{\nu}^2_j, F_{\boldsymbol{\theta}}) + \psi\left(W^2_{k_j,j}(\boldsymbol{z}, F_{\boldsymbol{\theta}}), \nu^2_j\right) = 0.
\end{aligned}$$

For consistent estimators of the WV, we have that $\mathbb{E}_{F_{\boldsymbol{\theta}}}[\psi(W^2_{k_j,j}(F_{\boldsymbol{\theta}}), \nu^2_j)] = 0$. We finally get

$$\mathrm{IF}(\boldsymbol{z}, \hat{\nu}^2_j, F_{\boldsymbol{\theta}}) \propto -\mathbf{D}^{-1}\psi\left(W^2_{k_j,j}(\boldsymbol{z}), \boldsymbol{\theta}\right)$$

with

$$\mathbf{D} = \mathbb{E}_{F_{\boldsymbol{\theta}}}\left[\frac{\partial}{\partial W^2_{k_j,j}}\psi\left(W^2_{k_j,j}, \nu^2_j\right)\right]_{W^2_{k_j,j}=W^2_{k_j,j}(F_{\boldsymbol{\theta}})} + \mathbb{E}_{F_{\boldsymbol{\theta}}}\left[\frac{\partial}{\partial \nu^2_j}\psi\left(W^2_{k_j,j}(F_{\boldsymbol{\theta}}), \nu^2_j\right)\right].$$

Since $\mathbf{D}$ does not depend on the contamination mass $\boldsymbol{z}$, the IF of the estimator of the WV is bounded if $\psi(\cdot)$ is bounded, thus concluding the proof. □

### *C.2. WV Identifiability*

In this appendix we discuss the identifiability of the WV $\nu^2_j$ when using the Huber and Tukey biweight $\psi$-functions. We first define the Tukey biweight function with redescending weights (see Beaton and Tukey, 1974). The biweight

$\psi$-function delivers the following weights $\omega(\cdot)$

$$\omega_{[Bi]}(r_{k_J,J};\nu_J^2,c) = \begin{cases} \left(\left(\frac{r_{k_J,J}}{c}\right)^2 - 1\right)^2 & \text{if } |r_{k_J,J}| \leq c \\ 0 & \text{if } |r_{k_J,J}| > c \end{cases}$$

and, if one supposes the normality for the wavelet coefficients, then the correction term $a(\nu_J^2)$ is

$$\begin{aligned} a_{[Bi]}(c) &= \mathbb{E}_\Phi\left[\omega^2_{[Bi]}(r_{k_J,J};\nu_J^2,c)r^2_{k_J,J}\right] \\ &= \frac{1}{c^8}\mu_c^{10} - \frac{4}{c^6}\mu_c^8 + \frac{6}{c^4}\mu_c^6 - \frac{4}{c^2}\mu_c^4 + \mu_c^2 \end{aligned} \tag{C-4}$$

with $\mu_c^i$ being the $i$-th truncated moment under the standard normal distribution between $-c$ and $c$.

On the other hand, Huber's $\psi$-function has well known properties and has easily tractable derivatives when developing its asymptotic properties. Its weights are given by

$$\omega_{[Hub]}(r_{k_J,J};\nu_J^2,c) = \min\left(1; \frac{c}{r_{k_J,J}}\right). \tag{C-5}$$

When using these functions, it is necessary to understand if they deliver functions which enable to identify the unknown parameter $\nu_J^2$ and state that

$$\mathbb{E}\left[\psi\left(W_{k_J,J},\kappa^2\right)\right] = 0$$

if and only if $\kappa^2 = \nu_J^2$ (i.e. there is a unique solution for $\nu_J^2$). This condition is often called "global identifiability" and it is an essential condition to prove consistency and asymptotic normality of the estimator which is often assumed for simplicity (as, for example, in Mondal and Percival, 2012a). To verify global identifiability, let us set the condition below:

**(C4)'** The process $(W_{k_J,J})$ is a zero-mean Gaussian process with autocovariance sequence $(\varphi_W(h))$ such that $|\varphi_W(h)| = \mathcal{O}(\rho^k)$, $0 < \rho < 1$.

Condition **(C4)'** is always verified regarding the mean constraint since all stationary models deliver zero-mean wavelet coefficients $(W_{k_J,J})$ with finite WV $\nu_J^2$ and many non-stationary models with stationary backward differences can also respect this condition. This is the case, in the time series setting for example, for all stationary ARMA and various state-space models. However, the assumption of a Gaussian model for $(W_{k_J,J})$ issued from the previously mentioned models is a relatively strong one but (apart from the case where $(X_k)$ is itself Gaussian) it is a frequently assumed condition for the wavelet coefficients and, according to the type of process, could be a reasonable approximation due to the averaging nature of the filter. Considering these observations and defining $\gamma \equiv c(\kappa^2/\nu_J^2)^{1/2}$, we have the following theorem.

**Theorem C.1.** *Under Condition (C4)' we have that $\nu_J^2$ is identifiable using the Huber weight function and is identifiable using the Tukey biweight function for $\gamma > 3.5$.*

The condition of $\gamma > 3.5$ is very mild. Indeed, the result of Theorem C.1 implies that the equation $\mathbb{E}[\psi(W_{k_j,j}, \kappa^2)] = 0$ has the unique solution $\kappa^2 = \nu_j^2$ if $\kappa$ belongs to the set $\{x \in \mathbb{R} \,|\, {}^{7\nu_j}/_{2c} < x < \infty\}$ for the Tukey biweight function. In other words, the parameter $\nu_j^2$ is identifiable if $c > 3.5$ so that it belongs to the previously defined set. Using the results of Theorem 3.2, this condition is very reasonable as it is satisfied for any efficiency larger than approximately 2.5%, an efficiency which is already too low to make any sense in practice. The proofs of Theorem C.1 can be found below.

*Proof.* Let us start with Huber weight function and, for this, let

$$X = \begin{cases} r_{k_j,j}^2 & \text{for } |r_{k_j,j}| \leq c \\ c^2 & \text{for } |r_{k_j,j}| > c \end{cases}$$

with $r_{k_j,j} = W_{k_j,j}/\kappa$, $\kappa^2 \in \{x \in \mathbb{R} \,|\, 0 < x < \infty\}$ and let us consider the function $\mathbb{E}[\psi(W_{k_j,j}, \kappa^2)]$. For Huber weights we define $q(r_{k_j,j}, c) \equiv \mathbb{E}\left[X - a_\psi(c)\right]$ where $a_\psi(c)$ is a constant for a given $c$. For global identifiability we need to prove that $q(r_{k_j,j}, c)$ has a unique solution in $\nu_j^2$ and to do so we prove that its derivative is a strictly monotone function in $\kappa^2$. Indeed, we have by definition that $\mathbb{E}[\psi(W_{k_j,j}, \kappa^2)] = 0$ if $\kappa^2 = \nu_j^2$ and if the derivative of $q(r_{k_j,j}, c)$ is strictly monotone then the solution is unique. Let us denote $\mathbb{P}[X]$ as the probability of $X$, $\alpha \equiv r_{k_j,j}\,{}^{\kappa}/_{\nu_j}$ and $\gamma \equiv c\,{}^{\kappa}/_{\nu_j} > 0$, then we have that

$$\begin{aligned}
\mathbb{E}\left[X - a_\psi(c)\right] &= \mathbb{E}\left[X\right] - a_\psi(c) \\
&= \mathbb{E}\left[X \middle| |r_{k_j,j}| \leq c\right] \mathbb{P}\left[|r_{k_j,j}| \leq c\right] \\
&+ \mathbb{E}\left[X \middle| |r_{k_j,j}| > c\right] \mathbb{P}\left[|r_{k_j,j}| > c\right] - a_\psi(c) \\
&= \frac{\nu_j^2}{\kappa^2}\left\{\mathbb{E}\left[X\frac{\kappa^2}{\nu_j^2} \,\middle|\, |\alpha| \leq \gamma\right] \mathbb{P}\left[|\alpha| \leq \gamma\right]\right\} \\
&+ \frac{\nu_j^2}{\kappa^2}\left\{\mathbb{E}\left[X\frac{\kappa^2}{\nu_j^2} \,\middle|\, |\alpha| > \gamma\right] \mathbb{P}\left[|\alpha| > \gamma\right]\right\} - a_\psi(c).
\end{aligned}$$

Denoting $\Phi(\cdot)$ and $\phi(\cdot)$ as being the Gaussian distribution and density functions respectively, using the results of Dhrymes (2005) we have

$$\begin{aligned}
\mathbb{E}\left[X - a_\psi(c)\right] =& \frac{c^2}{\gamma^2}\left(2\Phi(\gamma) - 2\gamma\phi(\gamma) - 1\right) + 2c^2\left(1 - \Phi(\gamma)\right) - a_\psi(c) = \\
& c^2 \underbrace{\left(\frac{2\Phi(\gamma)}{\gamma^2} - \frac{2\phi(\gamma)}{\gamma} - \frac{1}{\gamma^2} - 2\Phi(\gamma)\right)}_{f(\gamma)} + 2c^2 - a_\psi(c).
\end{aligned}$$

We define $g(\gamma) \equiv \mathbb{E}\left[X - a_\psi(c)\right]$ which has a unique solution for $\gamma$ if $f(\gamma)$ has a unique solution for $\gamma$. Hence, we focus on $f(\gamma)$ and take its derivative to understand if it is a strictly monotone function

$$\frac{\partial}{\partial\gamma}f(\gamma) = \frac{2(1-2\Phi(\gamma))}{\gamma^3} + \frac{4\phi(\gamma)}{\gamma^2} - \frac{2\phi'(\gamma)}{\gamma} - 2\phi(\gamma)$$

where $\phi'(\gamma) = \frac{\partial}{\partial\gamma}\phi(\gamma) = -\gamma\phi(\gamma)$ which finally gives us

$$\frac{\partial}{\partial\gamma}f(\gamma) = \frac{2}{\gamma^3}\underbrace{(2\gamma\phi(\gamma)+1-2\Phi(\gamma))}_{\text{A}}. \tag{C-6}$$

If we prove that the term A in (C-6) is strictly positive or negative, we prove that the derivative is too. By rewriting A we have

$$2\gamma\phi(\gamma)+1-2\Phi(\gamma) = 2\gamma\phi(\gamma)+2\Phi(0)-2\Phi(\gamma) = 2(\gamma\phi(\gamma)+\Phi(0)-\Phi(\gamma))$$

and we prove that this is quantity is strictly negative since $\gamma\phi(\gamma) < \Phi(\gamma) - \Phi(0)$ given that $\gamma > 0$.

Now, let us prove the identifiability for the Tukey biweight function. Therefore, in the same manner let

$$X = \begin{cases} \left(\left(\frac{r_{k_\jmath,\jmath}}{c}\right)^2 - 1\right)^4 r^2_{k_\jmath,\jmath} & \text{for } |r_{k_\jmath,\jmath}| \leq c \\ 0 & \text{for } |r_{k_\jmath,\jmath}| > c \end{cases}$$

and let $\kappa$ belong to the set $\{x \in \mathbb{R} \,|\, c^* < x < \infty\}$ where $c^*$ denotes a positive constant such that $c^* < \nu_\jmath$. Let us again follow the same procedure and notations as used for the proof of global identifiability of the Huber weights. With $a_\psi(c)$ being this time the correction term for the Tukey biweight function, in this case

we have

$$\mathbb{E}\left[X - a_\psi(c)\right] = \mathbb{E}\left[X\right] - a_\psi(c) = \mathbb{E}\left[X \middle| |r_{k_\jmath,\jmath}| \le c\right] \mathbb{P}\left[|r_{k_\jmath,\jmath}| \le c\right] - a_\psi(c)$$

$$= \frac{\nu_\jmath^{10}}{\kappa^{10} c^8} \left\{ \mathbb{E}\left[ r_{k_\jmath,\jmath}^{10} \frac{\kappa^{10}}{\nu_\jmath^{10}} \;\middle|\; |\alpha| \le \gamma \right] \mathbb{P}\left[|\alpha| \le \gamma\right] \right\}$$

$$- \frac{4\nu_\jmath^{8}}{\kappa^{8} c^6} \left\{ \mathbb{E}\left[ r_{k_\jmath,\jmath}^{8} \frac{\kappa^{8}}{\nu_\jmath^{8}} \;\middle|\; |\alpha| \le \gamma \right] \mathbb{P}\left[|\alpha| \le \gamma\right] \right\}$$

$$+ \frac{6\nu_\jmath^{6}}{\kappa^{6} c^4} \left\{ \mathbb{E}\left[ r_{k_\jmath,\jmath}^{6} \frac{\kappa^{6}}{\nu_\jmath^{6}} \;\middle|\; |\alpha| \le \gamma \right] \mathbb{P}\left[|\alpha| \le \gamma\right] \right\}$$

$$- \frac{4\nu_\jmath^{4}}{\kappa^{4} c^2} \left\{ \mathbb{E}\left[ r_{k_\jmath,\jmath}^{4} \frac{\kappa^{4}}{\nu_\jmath^{4}} \;\middle|\; |\alpha| \le \gamma \right] \mathbb{P}\left[|\alpha| \le \gamma\right] \right\}$$

$$+ \frac{\nu_\jmath}{\kappa} \left\{ \mathbb{E}\left[ r_{k_\jmath,\jmath} \frac{\kappa}{\nu_\jmath} \;\middle|\; |\alpha| \le \gamma \right] \mathbb{P}\left[|\alpha| \le \gamma\right] \right\} - a_\psi(c)$$

$$= c^2 \Bigg[ \frac{1}{\gamma^{10}} \underbrace{\left(1890\Phi(\gamma) - \overbrace{2\gamma(945 + 315\gamma^2 + 63\gamma^4 + 9\gamma^6 + \gamma^8)}^{\mu_{10}^*} \phi(\gamma) - 945\right)}_{\mu_{10}}$$

$$- \frac{4}{\gamma^8} \underbrace{\left(210\Phi(\gamma) - \overbrace{2\gamma(105 + 35\gamma^2 + 7\gamma^4 + \gamma^6)}^{\mu_8^*} \phi(\gamma) - 105\right)}_{\mu_8}$$

$$+ \frac{6}{\gamma^6} \underbrace{\left(30\Phi(\gamma) - \overbrace{2\gamma(15 + 5\gamma^2 + \gamma^4)}^{\mu_6^*} \phi(\gamma) - 15\right)}_{\mu_6}$$

$$+ \frac{4}{\gamma^4} \underbrace{\left(6\Phi(\gamma) - \overbrace{2\gamma(3 + \gamma^2)}^{\mu_4^*} \phi(\gamma) - 3\right)}_{\mu_4} + \frac{1}{\gamma^2} \underbrace{\left(2\Phi(\gamma) - 2\gamma\phi(\gamma) - 1\right)}_{\mu_2} \Bigg] - a_\psi(c).$$

Next, we define $g(\gamma) \equiv \mathbb{E}\left[X - a_\psi(c)\right]$ and we know that $g(\gamma)$ has a unique solution in $\gamma$ if the expression in square brackets in $g(\gamma)$ has a unique solution in $\gamma$. Hence, by taking the derivative we obtain

$$\frac{\partial}{\partial\gamma} g(\gamma) = -\frac{\mu_{10}}{\gamma^{11}} + \frac{32}{\gamma^9}\mu_8 + \frac{1}{\gamma^{10}} \big(1890\phi(\gamma)$$

$$- \left(\left(1890 + 1890\gamma^2 + 630\gamma^4 + 126\gamma^6 + 18\gamma^8\right) \phi(\gamma) - \gamma\phi(\gamma)\mu_{10}^*\right) \big)$$

$$- \frac{4}{\gamma^8} \left(210\phi(\gamma) - \left(\left(210 + 210\gamma^2 + 70\gamma^4 + 14\gamma^6\right) \phi(\gamma) - \gamma\phi(\gamma)\mu_8^*\right)\right)$$

$$- \frac{36}{\gamma^7}\mu_6 + \frac{6}{\gamma^6} \left(30\phi(\gamma) - \left(\left(30 + 30\gamma^2 + 10\gamma^4\right) \phi(\gamma) - \gamma\phi(\gamma)\mu_6^*\right)\right)$$

$$+ \frac{16}{\gamma^5}\mu_4 - \frac{4}{\gamma^4} \left(6\phi(\gamma) - \left(\left(6 + 6\gamma^2\right) \phi(\gamma) - \gamma\phi(\gamma)\mu_4^*\right)\right)$$

$$- \frac{2}{\gamma^3}\mu_2 + \frac{1}{\gamma^2} \left(2\phi(\gamma) - \left(2\phi(\gamma) - 2\gamma^2\phi(\gamma)\right)\right)$$

whose value is strictly negative for $\gamma > 3.5$ thereby defining the value $c^* =$

$^{7\nu_j}/_{2c}$. □

### C.3. Scale-wise Asymptotic Properties of $\hat{\nu}_j^2$

In this appendix we discuss the asymptotic normality of the proposed M-estimator for each wavelet decomposition scale $j$. Before proving Theorem 3.1, we need the result of the following lemma if we consider using the Huber $\psi$-function.

**Lemma C.1.** *The function $\psi(W_{k_j,j}, \nu_j^2)$ using Huber weights is Bouligand-differentiable as follows*

$$\psi'(W_{k_j,j}, \nu_j^2) = \begin{cases} -\frac{W^2_{k_j,j}}{\nu_j^4} & \text{if } |r_{k_j,j}| \le c \\ 0 & \text{if } |r_{k_j,j}| > c \end{cases}$$

The proof of this lemma is given below.

*Proof.* Let us define $r_0 \equiv {}^{W_{k_j,j}}/\sqrt{\nu_0^2}$ and $r \equiv {}^{W_{k_j,j}}/\sqrt{\nu^2}$ where $\nu^2 = \nu_0^2 + h$. By the definition in Scholtes (2012), a function $f(\cdot)$ is Bouligand differentiable (B-differentiable) at point $x_0$ if it is directionally differentiable at this point and there exists a function $f'(\cdot)$ such that $f(x_0 + h) = f(x_0) + f'(x_0)h + o(h)$. Using the approach of Christmann and Messem (2008), we first show that the function $\psi(W_{k_j,j}, \nu_j^2)$ is first degree B-differentiable using Huber weights. Below are the computations of the B-derivatives for the five cases of the Huber weight function:

1. Setting $r_0 = c$ we have:
   - If $h \ge 0$ $(r \le c)$:

$$\begin{aligned} \psi'(W_{k_j,j}, \nu_0^2)(h) + o(h) &= \psi(W_{k_j,j}, \nu_0^2 + h) - \psi(W_{k_j,j}, \nu_0^2) \\ &= r^2 - a_\psi(c) - r_0^2 + a_\psi(c) \\ &= \frac{W^2_{k_j,j}}{\nu_0^2 + h} - \frac{W^2_{k_j,j}}{\nu_0^2} \\ &= \frac{W^2_{k_j,j}}{\nu_0^2}\left(\frac{-h}{\nu_0^2 + h}\right) \\ &= -\frac{W^2_{k_j,j}}{\nu_0^2}\left(\frac{h}{\nu_0^2} - \frac{h^2}{\nu_0^2(\nu_0^2 + h)}\right) \\ &= -\frac{W^2_{k_j,j}}{\nu_0^4}h + \underbrace{\frac{W^2_{k_j,j}h^2}{\nu_0^2(\nu_0^2 + h)}}_{o(h)} \equiv \Delta \end{aligned}$$

   - If $h < 0$ $(r > c)$:

$$\begin{aligned} \psi'(W_{k_j,j}, \nu_0^2)(h) + o(h) &= \psi(W_{k_j,j}, \nu_0^2 + h) - \psi(W_{k_j,j}, \nu_0^2) \\ &= c^2 - a_\psi(c) - r_0^2 + a_\psi(c) = c^2 - c^2 = 0 \end{aligned}$$

2. Setting $r_0 = -c$ we have:
   - If $h < 0$ ($r < -c$):

$$\begin{aligned}\psi'(W_{k_\jmath,\jmath}, \nu_0^2)(h) + o(h) &= \psi(W_{k_\jmath,\jmath}, \nu_0^2 + h) - \psi(W_{k_\jmath,\jmath}, \nu_0^2) \\ &= c^2 - a_\psi(c) - r_0^2 + a_\psi(c) = 0\end{aligned}$$

   - If $h \geq 0$ ($r \geq -c$):

$$\begin{aligned}\psi'(W_{k_\jmath,\jmath}, \nu_0^2)(h) + o(h) &= \psi(W_{k_\jmath,\jmath}, \nu_0^2 + h) - \psi(W_{k_\jmath,\jmath}, \nu_0^2) \\ &= r^2 - a_\psi(c) - r_0^2 + a_\psi(c) \\ &= \ldots = \Delta\end{aligned}$$

3. Setting $r_0 > c$ we have:

$$\begin{aligned}\psi'(W_{k_\jmath,\jmath}, \nu_0^2)(h) + o(h) &= \psi(W_{k_\jmath,\jmath}, \nu_0^2 + h) - \psi(W_{k_\jmath,\jmath}, \nu_0^2) \\ &= c^2 - a_\psi(c) - r_0^2 + a_\psi(c) = 0\end{aligned}$$

4. Setting $r_0 < -c$ we have:

$$\begin{aligned}\psi'(W_{k_\jmath,\jmath}, \nu_0^2)(h) + o(h) &= \psi(W_{k_\jmath,\jmath}, \nu_0^2 + h) - \psi(W_{k_\jmath,\jmath}, \nu_0^2) \\ &= c^2 - a_\psi(c) - r_0^2 + a_\psi(c) = 0\end{aligned}$$

5. Setting $-c < r_0 < c$ we have:

$$\begin{aligned}\psi'(W_{k_\jmath,\jmath}, \nu_0^2)(h) + o(h) &= \psi(W_{k_\jmath,\jmath}, \nu_0^2 + h) - \psi(W_{k_\jmath,\jmath}, \nu_0^2) \\ &= r^2 - a_\psi(c) - r_0^2 + a_\psi(c) \\ &= \ldots = \Delta\end{aligned}$$

We therefore have that the first B-derivative of the function $\psi(W_{k_\jmath,\jmath}, \nu_\jmath^2)$ is given by

$$\psi'(W_{k_\jmath,\jmath}, \nu_\jmath^2) = \begin{cases} -\frac{W_{k_\jmath,\jmath}^2}{\nu_\jmath^4} & \text{if } |r_{k_\jmath,\jmath}| \leq c \\ 0 & \text{if } |r_{k_\jmath,\jmath}| > c \end{cases}$$

The approach used in this proof can be used to obtain expressions for the B-derivatives of other piecewise differentiable weight functions (see Scholtes, 2012). It can be seen how it extends the classic derivative for $|r_0| < c$ also to the points $\nu_0^2$ such that $|r_0| = c$. However, the Frechet differentiability of this function has also been discussed in Clarke (1986).

□

Technical Lemma C.1 is useful for the results on asymptotic normality of the proposed estimator to hold in case the choice of the $\psi$-function corresponds to the Huber $\psi$-function, proving that this function respects Condition **(C6)** which is required by Theorem 3.1. The proof of this theorem, which is valid for a general $\psi$-function that respects Condition **(C6)**, is given below.

*Proof.* Since $(W_{k_\jmath,\jmath})$ is a strictly stationary ergodic process, then so is $\big(\psi(W_{k_\jmath,\jmath},\nu_\jmath^2)\big)$ which is a bounded and time-invariant function of it. Hence, $\big(\psi(W_{k_\jmath,\jmath},\nu_\jmath^2)\big)$ satisfies the Uniform Weak Law of Large Numbers (UWLLN) under Theorem 4.1 of Wooldridge (1994). Condition **(C2)** implies that there is a unique minimum for the function $q(W_{k_\jmath,\jmath},\nu_\jmath^2) = -\int \psi(W_{k_\jmath,\jmath},\nu_\jmath^2)\mathrm{d}\nu_\jmath^2$ and, with condition **(C1)**, the weak consistency of $M$-estimators (Theorem 4.3, Wooldridge, 1994) yields

$$\hat{\nu}_\jmath^2 \xrightarrow{\mathcal{P}} \nu_\jmath^2.$$

Given condition **(C6)**, let us denote $\psi'(W_{k_\jmath,\jmath},{\nu_\jmath}^2) = \partial/\partial{\nu_\jmath}^2\, \psi(W_{k_\jmath,\jmath},{\nu_\jmath}^2)$ and apply a Maclaurin expansion of $\sum_{k_\jmath\in\mathcal{K}_\mathcal{J}} \psi(W_{k_\jmath,\jmath},\hat{\nu}_\jmath^2)$ around $\nu_\jmath^2$ obtaining

$$\sum_{k_\jmath\in\mathcal{K}_\mathcal{J}} \psi(W_{k_\jmath,\jmath},\hat{\nu}_\jmath^2) = \sum_{k_\jmath\in\mathcal{K}_\mathcal{J}} \psi(W_{k_\jmath,\jmath},\nu_\jmath^2) + \sum_{k_\jmath\in\mathcal{K}_\mathcal{J}} \psi'(W_{k_\jmath,\jmath},{\nu_\jmath^*}^2)(\hat{\nu}_\jmath^2-\nu_\jmath^2) = 0$$

where

$$\|{\nu_\jmath^*}^2 - \nu_\jmath^2\| \leq \|\hat{\nu}_\jmath^2 - \nu_\jmath^2\|. \tag{C-7}$$

Multiplying by $\sqrt{M_\jmath}$ and rewriting yields

$$\sqrt{M_\jmath}(\hat{\nu}_\jmath^2-\nu_\jmath^2) = \Bigg[\underbrace{-\frac{1}{M_\jmath}\sum_{k_\jmath\in\mathcal{K}_\mathcal{J}} \psi'(W_{k_\jmath,\jmath},{\nu_\jmath^*}^2)}_{\mathrm{A}_\jmath}\Bigg]^{-1} \underbrace{\frac{1}{\sqrt{M_\jmath}}\sum_{k_\jmath\in\mathcal{K}_\mathcal{J}} \psi(W_{k_\jmath,\jmath},\nu_\jmath^2)}_{\mathrm{B}_\jmath}.$$

Let us start from term $\mathrm{A}_\jmath$. We can rewrite this term as

$$-\frac{1}{M_\jmath}\sum_{k_\jmath\in\mathcal{K}_\mathcal{J}} \psi'(W_{k_\jmath,\jmath},{\nu_\jmath^*}^2) = -\frac{1}{M_\jmath}\sum_{k_\jmath\in\mathcal{K}_\mathcal{J}} \psi'(W_{k_\jmath,\jmath},\nu_\jmath^2) - \frac{1}{M_\jmath}\sum_{k_\jmath\in\mathcal{K}_\mathcal{J}} \underbrace{[\psi'(W_{k_\jmath,\jmath},{\nu_\jmath^*}^2) - \psi'(W_{k_\jmath,\jmath},\nu_\jmath^2)]}_{\mathrm{C}_\jmath}.$$

Since $\big(\psi'(W_{k_\jmath,\jmath},\nu_\jmath^2)\big)$ is a time-invariant function of $\big(W_{k_\jmath,\jmath}\big)$, it is also a stationary and ergodic process (see Wooldridge, 1994). Let us define $m_\jmath = \mathbb{E}[-\psi'(W_{k_\jmath,\jmath},\nu_\jmath^2)]$, then by Birkhoff's ergodic theorem we know that

$$\frac{1}{M_\jmath}\sum_{k_\jmath\in\mathcal{K}_\mathcal{J}} -\psi'(W_{k_\jmath,\jmath},\nu_\jmath^2) \xrightarrow{a.s.} \mathbb{E}[-\psi'(W_{k_\jmath,\jmath},\nu_\jmath^2)] = m_\jmath.$$

As for term $\mathrm{C}_\jmath$, it is also a stationary and ergodic process. Since $\hat{\nu}_\jmath^2$ is a consistent estimator of $\nu_\jmath^2$, by (C-7) so is ${\nu_\jmath^*}^2$ which yields $\mathbb{E}[\mathrm{C}_\jmath] = 0$ for $M_\jmath \to \infty$.

Hence, by again using Birkhoff's ergodic theorem, we have that

$$\frac{1}{M_\jmath}\sum_{k_\jmath\in\mathcal{K}_\mathcal{J}} [\psi'(W_{k_\jmath,\jmath},{\nu_\jmath^*}^2) - \psi'(W_{k_\jmath,\jmath},\nu_\jmath^2)] \xrightarrow{a.s.} 0$$

which finally yields

$$\mathrm{A}_\jmath \xrightarrow{a.s.} m_\jmath.$$

Let us now focus on term $\mathrm{B}_\jmath$ and let us define $S_{M_\jmath} = \sum_{k_\jmath \in \mathcal{K}_\mathcal{J}} \psi(W_{k_\jmath,\jmath}, \nu_\jmath^2)$, with $\left(\psi(W_{k_\jmath,\jmath}, \nu_\jmath^2)\right)$ being a stationary ergodic process with $\mathbb{E}[S_{M_\jmath}] = 0$ by definition and $\sigma^2_{M_\jmath} = Var[S_{M_\jmath}] \to \infty$. Then, following Theorem 3 of Denker (1986), we have

$$\frac{S_{M_\jmath}}{\sigma_{M_\jmath}} \xmapsto[M_\jmath \to \infty]{\mathcal{D}} \mathcal{N}(0,1)$$

if $S^2_{M_\jmath}/\sigma^2_{M_\jmath}$ is a uniformly integrable sequence. To show the latter, we use the criterion of Billingsley (2009) for uniformly integrable sequences based on which we need to prove that there exists a $\delta > 0$ such that

$$\sup_{M_\jmath} \mathbb{E}\left[\left|\frac{S^2_{M_\jmath}}{\sigma^2_{M_\jmath}}\right|^{1+\delta}\right] < \infty.$$

Let us take $\delta = 1$ so that we have

$$\sup_{M_\jmath} \mathbb{E}\left[\frac{S^4_{M_\jmath}}{\sigma^4_{M_\jmath}}\right] < \infty. \tag{C-8}$$

Let us define $Z_{k_\jmath} = \psi(W_{k_\jmath,\jmath}, \nu_\jmath^2)$ and let $Var[Z_{k_\jmath}] = S_\psi(0) < \infty$ (i.e. the power spectral density of $\left(\psi(W_{k_\jmath,\jmath}, \nu_\jmath^2)\right)$ at zero frequency is finite given that $\psi(\cdot)$ is a function of bounded variation). Let us consider the indices $i, j, k, l = 1, \dots, M_\jmath$, representing four distinct points on the $D$-dimensional wavelet decomposition lattice. We know that the fourth and lower order moments of $Z_{k_\jmath}$ are finite based on condition **(C5)** since they are bounded functions, hence $|\mathbb{E}[Z_i^4]|$ is also finite and is bounded by a quantity we denote as $T$. By the Cauchy-Schwarz inequality we have that also $|\mathbb{E}[Z_i^3 Z_j]|$, $|\mathbb{E}[Z_i^2 Z_j^2]|$, $|\mathbb{E}[Z_i^2 Z_j Z_k]|$ and $|\mathbb{E}[Z_i Z_j Z_k Z_l]|$ are bounded by $T$. Considering the above bound $T$ and defining $a_{k_\jmath}$ as the exponent for the $k_\jmath^{th}$ term in the expansion ($a_{k_\jmath} = 0, \dots, 4, a_{k_\jmath} \in \mathbb{N}$), using the multinomial theorem we have

$$\begin{aligned}
\mathbb{E}[S^4_{M_\jmath}] =& \mathbb{E}\left[\left(\sum_{k_\jmath \in \mathcal{K}_\jmath} Z_{k_\jmath}\right)^4\right] = \mathbb{E}\left[\sum_{a_1+\ldots+a_{M_\jmath}=4} \underbrace{\binom{4}{a_1, \ldots, a_{M_\jmath}}}_{\leq 4!} Z_1^{a_1} \cdots Z_{M_\jmath}^{a_{M_\jmath}}\right] \\
\leq& \mathbb{E}\left[\sum_{a_1+\ldots+a_{M_\jmath}=4} 4!\ Z_1^{a_1} \cdots Z_{M_\jmath}^{a_{M_\jmath}}\right] \\
=& \sum_{a_1+\ldots+a_T=4} 4!\ \underbrace{\mathbb{E}\left[Z_1^{a_1} \cdots Z_{M_\jmath}^{a_{M_\jmath}}\right]}_{\leq T} \leq \underbrace{\sum_{a_1+\ldots+a_{M_\jmath}=4}}_{\mathcal{O}(M_\jmath^4)} 4!\ T.
\end{aligned}$$

Now let us define $C(M_\jmath) = \frac{2}{M_\jmath(M_\jmath - 1)} \sum_{k_\jmath \in \mathcal{K}_\jmath} \sum_{\forall k'_\jmath < k_\jmath} \rho_{k_\jmath, k'_\jmath}$, with $\rho_{k_\jmath, k'_\jmath}$ being the correlation between $Z_{k_\jmath}$ and $Z'_{k_\jmath}$ and $0 \leq |C(M_\jmath)| \leq 1, \forall M_\jmath$. We then have

$$
\begin{aligned}
Var[S_{M_\jmath}]^2 =& Var\left[\sum_{k_\jmath \in \mathcal{K}_\jmath} Z_{k_\jmath}\right]^2 = \left(\sum_i Var[Z_i] + 2\sum_i \sum_{j<i} Cov[Z_i, Z_j]\right)^2 \\
&= \left(\sum_i Var[Z_i] + 2\sum_i \sum_{j<i} \rho_{i,j} Var[Z_i]\right)^2 \\
&= \left(\sum_i Var[Z_i] + 2\sum_i \sum_{j<i} C(M_\jmath) Var[Z_i]\right)^2 \\
&= (M_\jmath Var[Z_i] + 2M_\jmath(M_\jmath - 1)C(M_\jmath)Var[Z_i])^2 \\
&= M_\jmath^2 Var[Z_i]^2 + 4\underbrace{M_\jmath^2(M_\jmath - 1)}_{\mathcal{O}(M_\jmath^3)} Var[Z_i]^2 C(M_\jmath) + \\
&4\underbrace{M_\jmath^2(M_\jmath - 1)^2}_{\mathcal{O}(M_\jmath^4)} C(M_\jmath)^2 Var[Z_i]^2.
\end{aligned}
$$

Considering Condition (C-8) and the order of the expressions in $T$, we therefore have

$$
\sup_{M_\jmath} \frac{M_\jmath^4 24T}{M_\jmath^2 Var[Z_{k_\jmath}]^2 + 4M_\jmath^3 Var[Z_{k_\jmath}]C(M_\jmath) + 4M_\jmath^4 C(M_\jmath)^2 Var[Z_{k_\jmath}]^2} < \infty.
$$

Hence we have that $\frac{S^2_{M_\jmath}}{\sigma^2_{M_\jmath}}$ is uniformly integrable, thereby giving us

$$
\mathrm{B}_\jmath = \frac{1}{\sqrt{M_\jmath h(M_\jmath)}} \sum_{k_\jmath \in \mathcal{K}_\jmath} \psi(W_{k_\jmath,\jmath}, \nu_\jmath^2) \xrightarrow[M_\jmath \to \infty]{\mathcal{D}} \mathcal{N}\left(0, S_\psi(0)\right).
$$

where $h(\cdot)$ is a slowly varying function. Using Condition **(C4)** to fulfill the assumptions of Corollary 5.1 in Hall and Heyde (2014) for term $\mathrm{B}_\jmath$ and applying Slutsky's theorem on both $\mathrm{A}_\jmath$ and $\mathrm{B}_\jmath$ we finally obtain

$$
\sqrt{M_\jmath}\left(\hat{\nu}_\jmath^2 - \nu_\jmath^2\right) \xrightarrow[M_\jmath \to \infty]{\mathcal{D}} \mathcal{N}\left(0, \frac{S_\psi(0)}{m_\jmath^2}\right). \tag{C-9}
$$

□

### C.4. *Proof of Condition (C7) for Equal Number of Observations per Dimension*

*Proof.* If $K_d = K$, $\forall d$, then we have that $J_d = J$, $\forall d$, $a_d = a$, $\forall d$ and Condition **(C7)** therefore becomes

$$\frac{(\lfloor \log_2(K) \rfloor - a)^D}{\sqrt{(K - L_J + 1)^D}}.$$

By the definition of $J$, we know that $L_J = K/2$ and that

$$a = \log_2\left(\frac{2K(L_1 - 1)}{K + 2L_1 - 4}\right) = \log_2\left(\frac{2(L_1 - 1)}{1 + \frac{2L_1 - 4}{K}}\right) > 0$$

since $K \geq 2L_1$ and $L_1 \geq 2$ necessarily. We therefore have that Condition **(C7)** is bounded from above as follows

$$\frac{(\lfloor \log_2(K) \rfloor - a)^D}{\sqrt{(K - L_J + 1)^D}} \propto \frac{(\log_2(K) - a)^D}{\sqrt{(K - \frac{K}{2} + 1)^D}} < \frac{2^{D/2}(\log_2(K))^D}{K^{D/2}}.$$

Applying Hopital's rule $D$ times to the last expression gives us

$$\lim_{K \to \infty} \frac{(D-1)!}{K^{D/2} D^{(D-1)} \ln(2)^D 2^{D/2}} = 0$$

which implies that Condition **(C7)** is satisfied when the number of observations in each dimension are equal.

□

### C.5. *Mean Square Consistency of $\hat{\boldsymbol{\nu}}$*

Below we give the proof of Corollary 3.1.

*Proof.* Using the results of Proof C.3 we have that

$$\sqrt{M_J}\left(\hat{\nu}_J^2 - \nu_J^2\right) \xrightarrow{\mathcal{D}} \mathcal{N}\left(0, \frac{S_\psi(0)}{m_J^2}\right)$$

which by Condition **(C4)** implies

$$\left(\hat{\nu}_J^2 - \nu_J^2\right) = \mathcal{O}_p\left(\frac{1}{\sqrt{M_J}}\right).$$

Then for the estimator $\hat{\boldsymbol{\nu}}$ we have

$$\|\hat{\boldsymbol{\nu}} - \boldsymbol{\nu}\| = \sqrt{\sum_{J=1}^{J}(\hat{\nu}_J^2 - \nu_J^2)^2} \leq \sqrt{J \max_J(\hat{\nu}_J^2 - \nu_J^2)^2} = \mathcal{O}_p\left(\sqrt{\frac{J}{M_{J^*}}}\right)$$

which by Condition **(C7)** implies

$$\|\hat{\boldsymbol{\nu}} - \boldsymbol{\nu}\| \xrightarrow{\mathcal{P}} 0$$

thus concluding the proof. □

### C.6. Joint Asymptotic Normality of $\bar{\boldsymbol{\nu}}$

Let us now give two technical corollaries which are needed to prove Lemma C.2 further on. This lemma is then necessary to obtain the joint asymptotic normality of $\bar{\boldsymbol{\nu}}$ and, consequently, of $\hat{\boldsymbol{\nu}}$.

**Corollary C.1.** *Let $\bar{\nu}_j^2$ be the implicit solution in $\nu_j^2$ of*

$$\sum_{k_j \in \mathcal{K}_{j*}} \psi(W_{k_j,j}, \nu_j^2) = 0. \tag{C-10}$$

*Then, under the conditions of Theorem 3.1, we have that*

$$\sqrt{M_{j*}}(\bar{\nu}_j^2 - \nu_j^2) \underset{M_{j*}\to\infty}{\overset{\mathcal{D}}{\longmapsto}} \mathcal{N}\left(0, \bar{\sigma}^2\right)$$

*where $\bar{\sigma}^2 < \infty$ is a known variance, which implies*

$$(\bar{\nu}_j^2 - \nu_j^2) = \mathcal{O}_p\left(\frac{1}{\sqrt{M_{j*}}}\right).$$

**Corollary C.2.** *Let*

$$D_j = \sum_{k_j \in \mathcal{K}_{j*}} (\psi'(W_{k_j,j}, \nu_j^2) - m_j) \tag{C-11}$$

*with $m_j = \mathbb{E}[\psi'(W_{k_j,j}, \nu_j^2)]$ and defining $D_{j*} = {}^1/_{M_{j*}} D_j$. Then, under the conditions of Theorem 3.1 and Condition **(C8)**, we have that*

$$\frac{D_{j*}}{Var[D_{j*}]} \underset{M_j\to\infty}{\overset{\mathcal{D}}{\longmapsto}} \mathcal{N}\left(0, 1\right)$$

*which implies*

$$(D_{j*} - m_j) = \mathcal{O}_p\left(\frac{1}{\sqrt{M_{j*}}}\right).$$

*Proof.* The proof simply uses the results in Proof C.3. Indeed, $\bar{\nu}_j^2$ is another estimator which is different from $\hat{\nu}_j^2$ only because it is based on fewer observations (i.e. $M_{j*} \leq M_j$, $\forall j$) and therefore converges in distribution at a rate of $\mathcal{O}_p({}^1/\sqrt{M_{j*}}) \geq \mathcal{O}_p({}^1/\sqrt{M_j})$. As for term $D_j*$, this is a stationary and ergodic process since it is a time-invariant function of $(W_{k_j,j})$ and we can therefore define $Z_{k_j} = (\psi'(W_{k_j,j}, \nu_j^2) - m_j)$ as in Proof C.3. Under the same conditions and arguments, we therefore have that ${D_j^2}/{Var[D_j]^2}$ is a uniformly integrable sequence and we consequently have that $(D_{j*} - m_j)$ is of order $\mathcal{O}_p\left({}^1/\sqrt{M_{j*}}\right)$. □

Using these corollaries, we have the following lemma.

**Lemma C.2.** *Under the conditions of Corollaries C.1 and C.2, consider term*

$$A_{\jmath *} = \frac{1}{M_{\jmath *}} \sum_{k_\jmath \in \mathcal{K}_{\jmath *}} \psi'(W_{k_\jmath,\jmath}, \nu_\jmath^2)$$

*and* $m_\jmath = \mathbb{E}[\psi'(W_{k_\jmath,\jmath}, \nu_\jmath^2)]$. *Then we have that*

$$A_{\jmath *}^{-1} - m_\jmath^{-1} = \mathcal{O}_p\left(\frac{1}{\sqrt{M_{\jmath *}}}\right).$$

*Proof.* Assuming the conditions for Corollaries C.1 and C.2 hold and using the definition of $\bar{\nu}_\jmath^2$ given in (C-10), let us apply a Maclaurin expansion of $\sum_{k_\jmath \in \mathcal{K}_{\jmath *}} \psi(W_{k_\jmath,\jmath}, \bar{\nu}_\jmath^2)$ around $\nu_\jmath^2$

$$\sum_{k_\jmath \in \mathcal{K}_{\jmath *}} \psi(W_{k_\jmath,\jmath}, \bar{\nu}_\jmath^2) = \sum_{k_\jmath \in \mathcal{K}_{\jmath *}} \psi(W_{k_\jmath,\jmath}, \nu_\jmath^2) + \sum_{k_\jmath \in \mathcal{K}_{\jmath *}} \psi'(W_{k_\jmath,\jmath}, \nu_\jmath^{*2}) \left(\bar{\nu}_\jmath^2 - \nu_\jmath^2\right)$$

In addition, let us also apply a Taylor expansion of $\sum_{k_\jmath \in \mathcal{K}_{\jmath *}} \psi(W_{k,\jmath}, \bar{\nu}_\jmath^2)$ around $\nu_\jmath^2$

$$\begin{aligned}\sum_{k_\jmath \in \mathcal{K}_{\jmath *}} \psi(W_{k_\jmath,\jmath}, \bar{\nu}_\jmath^2) \quad = \quad & \sum_{k_\jmath \in \mathcal{K}_{\jmath *}} \psi(W_{k_\jmath,\jmath}, \nu_\jmath^2) + \sum_{k_\jmath \in \mathcal{K}_{\jmath *}} \psi'(W_{k_\jmath,\jmath}, \nu_\jmath^2) \left(\bar{\nu}_\jmath^2 - \nu_\jmath^2\right) + \\ & \sum_{k_\jmath \in \mathcal{K}_{\jmath *}} \gamma_{k_\jmath,\jmath}\end{aligned}$$

where $\gamma_{k_\jmath,\jmath} \equiv \mathcal{O}\left(\left(\bar{\nu}_\jmath^2 - \nu_\jmath^2\right)^2\right) = \mathcal{O}_p\left({}^1/_{M_{\jmath *}}\right)$ based on Corollary C.1.

These expansions imply that

$$\sum_{k_\jmath \in \mathcal{K}_{\jmath *}} \psi'(W_{k_\jmath,\jmath}, \nu_\jmath^{*2}) \left(\bar{\nu}_\jmath^2 - \nu_\jmath^2\right) = \sum_{k_\jmath \in \mathcal{K}_{\jmath *}} \psi'(W_{k_\jmath,\jmath}, \nu_\jmath^2) \left(\bar{\nu}_\jmath^2 - \nu_\jmath^2\right) + \sum_{k_\jmath \in \mathcal{K}_{\jmath *}} \gamma_{k_\jmath,\jmath}$$

which, by rearranging, delivers

$$\sum_{k_\jmath \in \mathcal{K}_{\jmath *}} \psi'(W_{k_\jmath,\jmath}, \nu_\jmath^{*2}) - \sum_{k_\jmath \in \mathcal{K}_{\jmath *}} \psi'(W_{k_\jmath,\jmath}, \nu_\jmath^2) = \sum_{k_\jmath \in \mathcal{K}_{\jmath *}} \frac{\gamma_{k_\jmath,\jmath}}{\left(\bar{\nu}_\jmath^2 - \nu_\jmath^2\right)}.$$

Since we have that ${}^{\gamma_{k_\jmath,\jmath}}/_{(\bar{\nu}_\jmath^2 - \nu_\jmath^2)} = \mathcal{O}_p\left({}^1/_{\sqrt{M_{\jmath *}}}\right)$, dividing by $M_{\jmath *}$ on both sides yields

$$\underbrace{\frac{1}{M_{\jmath *}} \sum_{k_\jmath \in \mathcal{K}_{\jmath *}} \psi'(W_{k_\jmath,\jmath}, \nu_\jmath^{*2})}_{A_{\jmath *}} - \underbrace{\frac{1}{M_{\jmath *}} \sum_{k_\jmath \in \mathcal{K}_{\jmath *}} \psi'(W_{k_\jmath,\jmath}, \nu_\jmath^2)}_{D_{\jmath *}} = \mathcal{O}_p\left(\frac{1}{\sqrt{M_{\jmath *}}}\right).$$

Using the results of Corollaries C.1 and C.2, we have

$$A_{\jmath*} - m_\jmath = \underbrace{A_{\jmath*} - D_{\jmath*}}_{\mathcal{O}_p(1/\sqrt{M_{\jmath*}})} + \underbrace{D_{\jmath*} - m_\jmath}_{\mathcal{O}_p(1/\sqrt{M_{\jmath*}})} = \mathcal{O}_p\left(\frac{1}{\sqrt{M_{\jmath*}}}\right)$$

which implies

$$A_{\jmath*}^{-1} - m_\jmath^{-1} = \mathcal{O}_p\left(\frac{1}{\sqrt{M_{\jmath*}}}\right).$$

□

Given the above lemma, below we give the proof of Theorem 3.2.

*Proof.* Let us define $U_\jmath \equiv \bar{\nu}_\jmath^2 - \nu_\jmath^2$, where $\bar{\nu}_\jmath^2$ is the solution for $\nu_\jmath^2$ in

$$\sum_{k_\jmath \in \mathcal{K}_{\jmath*}} \psi(W_{k_\jmath,\jmath}, \nu_\jmath^2) = 0.$$

From (C-8) we have that

$$\sqrt{M_{\jmath*}}\left(\bar{\nu}_\jmath^2 - \nu_\jmath^2\right) = \sqrt{M_{\jmath*}} U_\jmath = \mathrm{A}_{\jmath*}^{-1} \frac{1}{\sqrt{M_{\jmath*}}} \sum_{k_\jmath \in \mathcal{K}_{\jmath*}} \psi(W_{k_\jmath,\jmath}, \nu_\jmath^2)$$

where $\mathrm{A}_{\jmath*}$ is defined as in Lemma C.2. Now let $\sqrt{M_{\jmath*}} Q_\jmath \equiv m_\jmath^{-1} \frac{1}{\sqrt{M_{\jmath*}}} \sum_{k_\jmath \in \mathcal{K}_{\jmath*}} \psi(W_{k_\jmath,\jmath}, \nu_\jmath^2)$, let $\boldsymbol{a} = [a_\jmath]_{\jmath \in \mathcal{J}}$ be a vector of constants with at least one element different from zero and finally let $\boldsymbol{Q} = [Q_\jmath]_{\jmath \in \mathcal{J}}$. Then we have

$$\begin{aligned}
\sqrt{M_{\jmath*}} \boldsymbol{a}^T \boldsymbol{Q} &= \sum_{\jmath \in \mathcal{J}} a_\jmath m_\jmath^{-1} \frac{1}{\sqrt{M_{\jmath*}}} \sum_{k_\jmath \in \mathcal{K}_{\jmath*}} \psi(W_{k_\jmath,\jmath}, \nu_\jmath^2) \\
&= \frac{1}{\sqrt{M_{\jmath*}}} \sum_{k_\jmath \in \mathcal{K}_{\jmath*}} \sum_{\jmath \in \mathcal{J}} a_\jmath m_\jmath^{-1} \psi(W_{k_\jmath,\jmath}, \nu_\jmath^2) = \frac{1}{\sqrt{M_{\jmath*}}} \sum_{k_\jmath \in \mathcal{K}_{\jmath*}} R_{k_\jmath}
\end{aligned}$$

where $R_{k_\jmath} \equiv \sum_{\jmath \in \mathcal{J}} a_\jmath m_\jmath^{-1} \psi(W_{k_\jmath,\jmath}, \nu_\jmath^2)$. Being a time-invariant function of $(W_{k_\jmath,\jmath})$, we know that also $(R_{k_\jmath})$ is a stationary and ergodic process with $\mathbb{E}\left[\sum_{k_\jmath \in \mathcal{K}_{\jmath*}} R_{k_\jmath}\right] = 0$ and $\sigma_R^2 \equiv Var\left[\sum_{k_\jmath \in \mathcal{K}_{\jmath*}} R_{k_\jmath}\right] \xrightarrow{M_{\jmath*}} \infty$. By using the same argument as for $Z_{k_\jmath}$ in Appendix C.3, we have that $(\sum_{k_\jmath \in \mathcal{K}_{\jmath*}} R_{k_\jmath})^2/\sigma_R^2$ is a uniformly integrable sequence. Based on these results and again following Theorem 3 of Denker (1986) we have that

$$\sqrt{M_{\jmath*}} \boldsymbol{a}^T \boldsymbol{Q} \xrightarrow[M_\jmath \to \infty]{\mathcal{D}} \mathcal{N}\left(0, \sigma_Q^2\right).$$

Now let us take $\mathrm{B}_\jmath = \frac{1}{\sqrt{M_{\jmath*}}} \sum_{k_\jmath \in \mathcal{K}_{\jmath*}} \psi(W_{k_\jmath,\jmath}, \nu_\jmath^2)$ based on term B in (C-8) and consider

$$\sqrt{M_{\jmath*}} \sum_{\jmath \in \mathcal{J}} a_\jmath \left(U_\jmath - Q_\jmath\right) = \sum_{\jmath \in \mathcal{J}} a_\jmath \left(\sqrt{M_{\jmath*}} U_\jmath - \sqrt{M_{\jmath*}} Q_\jmath\right)$$

$$= \sum_{\jmath \in \mathcal{J}} a_\jmath \left(\mathrm{A}_{\jmath *}^{-1} - m_\jmath^{-1}\right) \mathrm{B}_\jmath .$$

Since from Lemma C.2 we have that $A_{\jmath *}^{-1} - m_\jmath^{-1} = \mathcal{O}_p({}^1/\sqrt{M_{\jmath *}})$ and $\mathrm{B}_\jmath$ converges to a Gaussian distribution, we have that

$$\sum_{\jmath \in \mathcal{J}} a_\jmath \left(\mathrm{A}_{\jmath *}^{-1} - m_\jmath^{-1}\right) \mathrm{B}_\jmath = \sum_{\jmath \in \mathcal{J}} \mathcal{O}_p\left(\frac{1}{\sqrt{M_{\jmath *}}}\right) = \mathcal{O}_p\left(\frac{J}{\sqrt{M_{\jmath *}}}\right)$$

where ${}^J/\sqrt{M_{\jmath *}} \to 0$ by Condition **(C7)**. Therefore we have that $\sqrt{M_{\jmath *}} \sum_{\jmath \in \mathcal{J}} a_\jmath \left(U_\jmath - Q_\jmath\right) = \mathcal{O}_p({}^J/\sqrt{M_{\jmath *}})$ and, by finally defining $\boldsymbol{U} = [U_\jmath]_{\jmath \in \mathcal{J}}$, we have

$$\begin{aligned}
\sqrt{M_{\jmath *}} \boldsymbol{a}^T \boldsymbol{U} &= \sqrt{M_{\jmath *}} \sum_{\jmath \in \mathcal{J}} a_\jmath \left(Q_\jmath + \left(U_\jmath - Q_\jmath\right)\right) \\
&= \sqrt{M_{\jmath *}} \boldsymbol{a}^T \boldsymbol{Q} + \mathcal{O}_p\left(\frac{J}{\sqrt{M_{\jmath *}}}\right) \xmapsto[M_{\jmath *} \to \infty]{\mathcal{D}} \mathcal{N}\left(0, \sigma_Q^2\right)
\end{aligned}$$

following Slutsky's theorem. Given these results and defining $\|\boldsymbol{s}\| = 1$, by the Cramer-Wold theorem we have

$$\sqrt{M_{\jmath *}} \boldsymbol{s}^T \boldsymbol{\Sigma}^{-1/2} \left(\bar{\boldsymbol{\nu}} - \boldsymbol{\nu}\right) \xmapsto[M_{\jmath *} \to \infty]{\mathcal{D}} \mathcal{N}\left(0, 1\right)$$

where, based on (C-9), it can be shown that $\boldsymbol{\Sigma} = \mathbf{M}^{-T} \mathbf{S}_\psi(\mathbf{0}) \mathbf{M}^{-1}$ with $\mathbf{S}_\psi(\mathbf{0})$ being the power spectral density of $\Psi(\mathbf{W}_{k_\jmath}, \boldsymbol{\nu})$ at zero-frequency and $\mathbf{M} = \mathbb{E}\left[-\frac{\partial}{\partial \boldsymbol{\nu}} \Psi(\mathbf{W}_{k_\jmath}, \boldsymbol{\nu})\right]$, thus concluding the proof. □

### *C.7. Joint Asymptotic Normality of $\hat{\boldsymbol{\nu}}$ for $D = 1$*

In this appendix we give Theorem C.2 and Corollary C.3 which are useful results to prove joint asymptotic properties of M-estimators when applied to different sample sizes which are related to each other (as is the case for wavelet decomposition or, for example, for autocovariance sequences at different lags). These results are useful to obtain Theorem 3.3 whose proof can be found at the end of this section. For this reason, let us state the following setting:

**(S5)** Let $M$ represent the number of vectorized coefficients $(W_{k_\jmath,\jmath})$ respecting conditions of Theorem 3.1 issued from a given filtering of a random field $(X_k)$ at scale $\jmath$, where the number of observations in the random field $N$ is such that $N > M$. Moreover, let $M^* < M$ and let $\hat{T}$ and $\bar{T}$ represent the same consistent and asymptotically normally distributed M-estimator, for the true parameter value $\theta_0$, computed on $M$ and $M^*$ observations respectively. Moreover, let the convergence of these estimators to a normal distribution be of order $\mathcal{O}_p({}^1/\sqrt{M})$ and $\mathcal{O}_p({}^1/\sqrt{M^*})$ for $\hat{T}$ and $\bar{T}$ respectively. Finally, let us define $\Delta = M - M^* > 0$ and assume $M > 2\Delta$, $\forall \Delta$, as well as $\Delta \underset{M \to \infty}{\longrightarrow} \infty$.

**Theorem C.2.** *Under Setting (S5) we have that*

$$(\hat{T} - \bar{T}) = \mathcal{O}_p\left(\frac{\sqrt{\Delta}}{M}\right).$$

*Proof.* Let $\hat{T}$ and $\bar{T}$ respectively be the implicit solutions, for a parameter $\theta$, of

$$\sum_{i=1}^{M} \psi(W_{i,\jmath}, \theta) = 0 \quad \text{and} \quad \sum_{i=1}^{M^*} \psi(W_{i,\jmath}, \theta) = 0.$$

Let us now make a Maclaurin expansion of $\sum_{i=1}^{M} \psi(W_{i,\jmath}, \hat{T})$ around $\bar{T}$ as follows

$$\sum_{i=1}^{M} \psi(W_{i,\jmath}, \hat{T}) = \sum_{i=1}^{M} \psi(W_{i,\jmath}, \bar{T}) + \sum_{i=1}^{M} \psi'(W_{i,\jmath}, T^*)(\hat{T} - \bar{T}) = 0$$

with $\|T^* - \bar{T}\| \leq \|\hat{T} - \bar{T}\|$. Rearranging we have

$$(\hat{T} - \bar{T}) = \left[-\sum_{i=1}^{M} \psi'(W_{i,\jmath}, T^*)\right]^{-1} \sum_{i=1}^{M} \psi(W_{i,\jmath}, \bar{T})$$

where

$$\sum_{i=1}^{M} \psi(W_{i,\jmath}, \bar{T}) = \underbrace{\sum_{i=1}^{M^*} \psi(W_{i,\jmath}, \bar{T})}_{=0} + \sum_{i=M-M^*+1}^{M} \psi(W_{i,\jmath}, \bar{T})$$

which leaves us with

$$(\hat{T} - \bar{T}) = \frac{\sqrt{\Delta}}{M} \left[\underbrace{-\frac{1}{M} \sum_{i=1}^{M} \psi'(W_{i,\jmath}, T^*)}_{A_\jmath}\right]^{-1} \underbrace{\frac{1}{\sqrt{\Delta}} \sum_{i=M-M^*+1}^{M} \psi(W_{i,\jmath}, \bar{T})}_{B_\jmath}.$$

Let us focus on term $B_\jmath$ and, with $b = \mathbb{E}[\psi(W_{i,\jmath}, \bar{T})]$, let us re-express it as

$$B_\jmath = \frac{1}{\sqrt{\Delta}} \sum_{i=M-M^*+1}^{M} \big(\psi(W_{i,\jmath}, \bar{T}) - b\big) + \sqrt{\Delta} b.$$

By using the same conditions and arguments as in Proof C.3, we have that

$$\frac{1}{\sqrt{\Delta}} \sum_{i=M-M^*+1}^{M} \big(\psi(W_{i,\jmath}, \bar{T}) - b\big) \xmapsto[\Delta \to \infty]{\mathcal{D}} \mathcal{N}\left(0, \sigma_\Delta^2\right)$$

for some $\sigma_\Delta^2 < \infty$ thereby implying that $1/\sqrt{\Delta} \sum_{i=M-M^*+1}^{M} \big(\psi(W_{i,\jmath}, \bar{T}) - b\big) = \mathcal{O}_p(1)$. Therefore, by Slutsky's theorem, $B_\jmath$ is normally distributed with mean $\sqrt{\Delta}b$. By taking a Taylor expansion of $b$ around $\theta_0$ and using Condition **(C6)** it is straightforward to verify that $b \to 0$ at the rate $\mathcal{O}_p(1/\sqrt{M^*})$ based on the fact that $(\bar{T} - \theta_0) = \mathcal{O}_p(1/\sqrt{M^*})$. This implies that

$$B_\jmath = \underbrace{\frac{1}{\sqrt{\Delta}} \sum_{i=M-M^*+1}^{M} \left(\psi(W_{i,\jmath}, \bar{T}) - b\right)}_{\mathcal{O}_p(1)} + \sqrt{\Delta} \underbrace{b}_{\mathcal{O}_p(1/\sqrt{M^*})} = \mathcal{O}_p\left(\frac{\sqrt{\Delta}}{\sqrt{M^*}}\right).$$

Using the results in Lemma C.2, we have that $A_\jmath = m_\jmath + \mathcal{O}_p(1/\sqrt{M})$ where $m_\jmath$ is a constant. Bringing everything together we have

$$(\hat{T} - \bar{T}) = \frac{\sqrt{\Delta}}{M} \left[ \underbrace{-\frac{1}{M} \sum_{i=1}^{M} \psi'(W_{i,\jmath}, T^*)}_{m_\jmath + \mathcal{O}_p(1/\sqrt{M})} \right]^{-1} \underbrace{\frac{1}{\sqrt{\Delta}} \sum_{i=M-M^*+1}^{M} \psi(W_{i,\jmath}, \bar{T})}_{\mathcal{O}_p(\sqrt{\Delta}/\sqrt{M^*})}.$$

This implies that when $M > 2\Delta$, we have that

$$(\hat{T} - \bar{T}) = \mathcal{O}_p\left(\frac{\sqrt{\Delta}}{M}\right).$$

□

**Corollary C.3.** *Let $\Delta = M - M^* > 0$ be fixed and let $\hat{T}$ and $\bar{T}$ be M-estimators as defined in Theorem C.2. Then we have that*

$$(\hat{T} - \bar{T}) = \mathcal{O}_p\left(\frac{1}{M}\right).$$

*Proof.* The proof follows directly from that of Theorem C.2. □

We can now give the proof of Theorem 3.3.

*Proof.* Let

$$\sqrt{M_{j*}} \boldsymbol{s}^T \boldsymbol{\Sigma}^{-1/2} \left(\hat{\boldsymbol{\nu}} - \boldsymbol{\nu}\right) = \sqrt{M_{j*}} \boldsymbol{s}^T \boldsymbol{\Sigma}^{-1/2} \left(\hat{\boldsymbol{\nu}} - \bar{\boldsymbol{\nu}}\right) + \sqrt{M_{j*}} \boldsymbol{s}^T \boldsymbol{\Sigma}^{-1/2} \left(\bar{\boldsymbol{\nu}} - \boldsymbol{\nu}\right)$$

where $\sqrt{M_{j*}} \boldsymbol{s}^T \boldsymbol{\Sigma}^{-1/2} \left(\bar{\boldsymbol{\nu}} - \boldsymbol{\nu}\right) \xmapsto[M_{j*}\to\infty]{\mathcal{D}} \mathcal{N}(0,1)$ by Theorem 3.2 and

$$\sqrt{M_{j*}} \boldsymbol{s}^T \boldsymbol{\Sigma}^{-1/2} \left(\hat{\boldsymbol{\nu}} - \bar{\boldsymbol{\nu}}\right) = \sqrt{M_{j*}} \sum_{j=1}^{J} a_j \left(\hat{\nu}_j^2 - \bar{\nu}_j^2\right)$$

where $a_j$ is an element of the constant $1 \times J$ vector $\boldsymbol{s}^T \boldsymbol{\Sigma}^{-1/2}$. If we take the order of this term, by Theorem C.2 we have that

$$\sqrt{M_{j*}} \boldsymbol{s}^T \boldsymbol{\Sigma}^{-1/2} \left(\hat{\boldsymbol{\nu}} - \bar{\boldsymbol{\nu}}\right) = \mathcal{O}_p\left(\sqrt{M_{j*}} \sum_{j=1}^{J} \frac{\sqrt{2^J - 2^j}}{M_j}\right)$$

with $M_j = N - 2^j + 1$. Let us first focus on the term in the sum of this order to obtain

$$\frac{\sqrt{2^J - 2^j}}{N - 2^j + 1} = \frac{\sqrt{2^J - 2^j}}{N} + \frac{\sqrt{2^J - 2^j}(2^j - 1)}{N M_j} < 2\frac{\sqrt{2^J - 2^j}}{N}$$

since ${}^{2^j-1}/_{M_j} < 1$. This leaves us with

$$\sqrt{M_{j*}} \sum_{j=1}^{J} \frac{\sqrt{2^J - 2^j}}{M_j} < 2\frac{\sqrt{M_{j*}}}{N} \sum_{j=1}^{J} \sqrt{2^J - 2^j}$$

and, by using $2^J - 2^j = 2^j(2^{J-j} - 1) < 2^j 2^J$, we have

$$2\frac{\sqrt{M_{j*}}}{N} \sum_{j=1}^{J} \sqrt{2^J - 2^j} < 2\frac{\sqrt{M_{j*}}}{N} 2^{J/2} \underbrace{\sum_{j=1}^{J} \sqrt{2^j}}_{=(2^{J/2}-1)(\sqrt{2}+2)} .$$

The order of the latter term is $\mathcal{O}_p(2^{J+1}\sqrt{M_{j*}}/N)$ and since $\sqrt{M_{j*}/N} < 1$, we finally have

$$2^{J+1}\frac{\sqrt{M_{j*}}}{N} < \frac{2^{J+1}}{\sqrt{N}} = 2\frac{N^{\alpha}}{\sqrt{N}}.$$

We conclude that

$$\sqrt{M_{j*}}\boldsymbol{s}^T\boldsymbol{\Sigma}^{-1/2}\left(\hat{\boldsymbol{\nu}} - \bar{\boldsymbol{\nu}}\right) = \mathcal{O}_p\left(N^{\alpha-1/2}\right)$$

which goes to zero since $0 < \alpha < 1/2$. By finally applying Slutsky's theorem, we obtain

$$\sqrt{M_{j*}}\boldsymbol{s}^T\boldsymbol{\Sigma}^{-1/2}\left(\hat{\boldsymbol{\nu}} - \boldsymbol{\nu}\right) \xrightarrow[M_{j*}\to\infty]{\mathcal{D}} \mathcal{N}(0,1).$$

□

## Appendix D: Cloud Images Application

The cloud data has already been analysed using the standard and robust estimators of WV respectively in Mondal and Percival (2012b) and Mondal and Percival (2012a). This data consists in four cloud regions taken from a larger temperature image made available by Geo-stationary Operational Environmental Satellites (GOES) Imagery. The study of cloud variability is paramount to understand the cloud-radiation interaction since this can generate three-dimensional radiative transfer effects which need to be taken into account for the modelling of global and regional climates (see, for example Davis et al., 2002). The image under study represents the stratocumulus clouds in the Chilean coast region which play a major role in the variations of temperatures and seasonal cycles in the east Pacific ocean since they reflect the light from the sun. In Mondal and Percival (2012b) they state that the four regions which were analyzed were picked by some atmospheric scientists who identified them as being of interest for the purpose of understanding the characteristics of these cloud formations whose dynamics are considered as being complex in this specific region (see Mondal and Percival, 2012b, for the details on these images). These studies are of interest also since the region is subject to atmospheric aerosol coming from industrial activities as well as Pockets of Open-Cells (POC) which are mainly responsible for the precipitations in the region.

Let us therefore reproduce the analysis made in Mondal and Percival (2012a) using the robust estimator $\hat{\boldsymbol{\nu}}$ proposed in this paper. The results are given in Figure 8 where we find the heat-maps of the cloud image itself (top row), the standard WV (top-middle row), the robust WV (bottom-middle row) and finally of the absolute difference between the standard and robust WV (bottom row). The heat-maps of the standard WV roughly correspond to the figures presented in Mondal and Percival (2012b) with some minor differences simply due to the different color-codes for representing the levels of the WV.

It can be observed how the first two regions have the same WV pattern both in the standard and robust case, as confirmed by the heat-maps of the absolute difference which are mostly white. This pattern is similar also for the third region in the standard case but is considerably different in the robust case where a high variability is detected mainly in the lower-left region instead of the upper-right and a slightly different variability pattern is detected also in the fourth region. If one considers what these four regions represent, this difference between standard and robust WV appears to be reasonable. Indeed, the first two regions represent "stable" regions, in the sense that the first region represents an already formed POC while the second one represents a uniform formation of stratocumulus clouds. For these two regions therefore, one would expect to well measure the cloud variability for these specific type of formations. However, the last two regions are characterized by "unstable" settings where the third region represents a formation of broken clouds while the fourth represents a POC that is in the process of being formed. These "instabilities" appear therefore to have an impact on the standard WV, assuming that the wavelet coefficients are roughly Gaussian, and although the data was preprocessed using a median filter, it would appear that a robust analysis should nevertheless be taken into consideration. The result of these analyses confirm the conclusions made in Mondal and Percival (2012a) and would consequently have an important impact on the choice of the climate models and the way cloud-radiation interaction is interpreted in these regions.

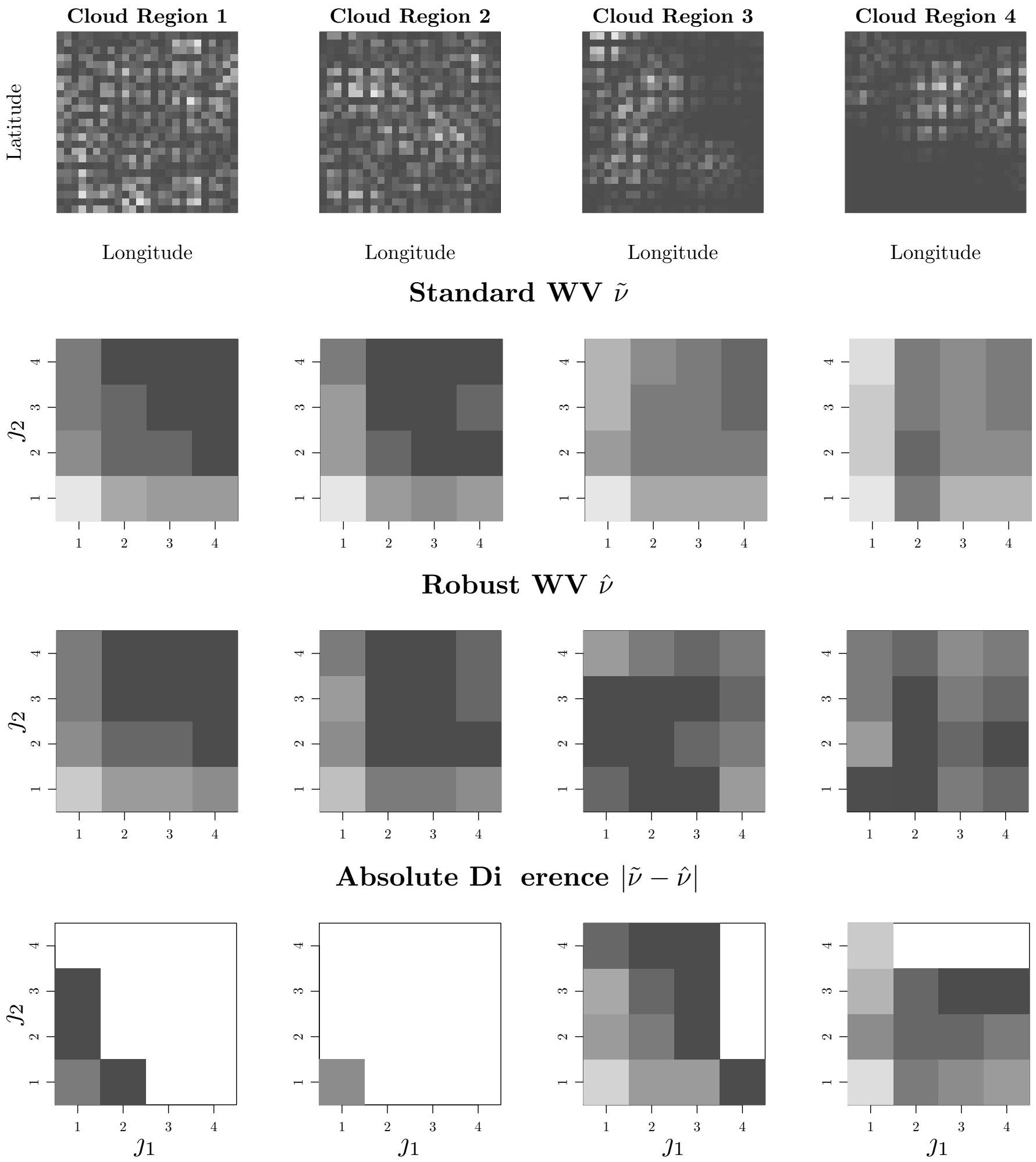

Fig 8: Heat-images of the raw cloud data for the four regions (top row), of the standard WV $\tilde{\boldsymbol{\nu}}$ (top-middle row), of the robust WV $\hat{\boldsymbol{\nu}}$ (bottom-middle row) and of their absolute difference (bottom row). The lighter shades indicate low values and darker shades indicate high values. The colour-coding is relative to each region (i.e. the WV has the same coding within the same region but not necessarily between regions).